\documentclass[conference]{IEEEtran}
\IEEEoverridecommandlockouts
\usepackage{cite}
\usepackage{amsmath,amssymb,amsfonts}
\usepackage{algorithmic}
\usepackage{graphicx}
\usepackage{textcomp}
\usepackage{xcolor}
\usepackage{siunitx}
\usepackage{listings}
\usepackage{soul}
\usepackage{xspace}
\usepackage{subcaption}
\usepackage{multicol}
\usepackage{float} 
\usepackage{etoolbox}

\usepackage{hyperref}
\hypersetup{breaklinks=true}

\definecolor{codegreen}{rgb}{0,0.6,0}
\definecolor{codegray}{rgb}{0.5,0.5,0.5}
\definecolor{codepurple}{rgb}{0.58,0,0.82}
\definecolor{backcolour}{rgb}{0.95,0.95,0.92}

\lstdefinestyle{mystyle}{
    backgroundcolor=\color{backcolour},   
    commentstyle=\color{codegreen},
    keywordstyle=\color{magenta},
    numberstyle=\tiny\color{codegray},
    stringstyle=\color{codepurple},
    basicstyle=\ttfamily\footnotesize,
    breakatwhitespace=false,         
    breaklines=true,                 
    captionpos=b,                    
    keepspaces=true,                 
    numbers=left,                    
    numbersep=5pt,                  
    showspaces=false,                
    showstringspaces=false,
    showtabs=false,                  
    tabsize=2
}

\def\BibTeX{{\rm B\kern-.05em{\sc i\kern-.025em b}\kern-.08em
    T\kern-.1667em\lower.7ex\hbox{E}\kern-.125emX}}
\begin{document}

\newcommand{\lm}[1]{\footnote{{\bf Luca: #1}}}
\newcommand{\ay}[1]{\footnote{{\bf Ahmed: #1}}}

\newcommand{\fakepar}[1]{\vspace{0mm}\noindent\textbf{#1.}}
\newcommand\code[1]{\textbf{\texttt{#1}}}
\newcommand\figref[1]{Fig.\,\ref{#1}}
\newcommand\secref[1]{Sec.\,\ref{#1}}
\newcommand\tabref[1]{Tab.\,\ref{#1}}
\newcommand\equationref[1]{Equation \ref{#1}}
\newcommand\appendixref[1]{Appendix \ref{#1}}

\newcommand{\deltainstruction}{$\delta$ instruction\xspace} 
\newcommand{\deltainstructions}{$\delta$ instructions\xspace} 
\newcommand{\reta}{\textsc{R\lowercase{e}TA}\xspace}
\newcommand{\deltatool}{\textsc{Delta}\xspace}
\newcommand{\aitreta}{\textsc{aiT-R\lowercase{e}TA}\xspace}
\newcommand{\ait}{\textsc{aiT}\xspace}

\makeatletter 
\newcommand{\linebreakand}{%
  \end{@IEEEauthorhalign}
  \hfill\mbox{}\par
  \mbox{}\hfill\begin{@IEEEauthorhalign}
}
\makeatother 

\title{Timing Analysis of Embedded Software Updates\\
}

\author{\IEEEauthorblockN{Ahmed El Yaacoub}
\IEEEauthorblockA{\textit{Uppsala University}\\ 
ahmed.el.yaacoub\textsuperscript{†}}
\and
\IEEEauthorblockN{Luca Mottola}
\IEEEauthorblockA{\textit{Politecnico di Milano  }\\ \textit{\& Uppsala University}\\
luca.mottola\textsuperscript{†}
}
\and
\IEEEauthorblockN{Thiemo Voigt}
\IEEEauthorblockA{\textit{Uppsala University} \\ \textit{\& RISE}\\
thiemo.voigt\textsuperscript{†}
}
\and
\IEEEauthorblockN{Philipp Rümmer}
\IEEEauthorblockA{\textit{University of Regensburg} \\ \textit{\& Uppsala University}\\
philipp.ruemmer\textsuperscript{*}
}
\linebreakand 
\IEEEauthorblockN{\textsuperscript{†}@angstrom.uu.se; \textsuperscript{*}@ur.de}
}

\maketitle

\newtoggle{extended}

\toggletrue{extended}

\begin{abstract}

  We present \reta (\underline{Re}lative \underline{T}iming \underline{A}nalysis), a \emph{differential timing analysis technique} to verify the impact of an update on the execution time of embedded software.
  Timing analysis is computationally expensive and labor intensive.
  Software updates render repeating the analysis from scratch a waste of resources and time, because their impact is \emph{inherently confined}.
  To determine this boundary, in \reta we apply a slicing procedure that identifies all relevant code segments and a statement categorization that determines how to analyze each such line of code.
    We adapt a subset of \reta for integration into aiT, an industrial timing analysis tool, and also develop a complete implementation in a tool called \deltatool.
    Based on staple benchmarks and realistic code updates from official repositories,  we test the accuracy by analyzing the worst-case execution time (WCET) before and after an update, comparing the measures with the use of  the unmodified aiT as well as real executions on embedded hardware.
  \deltatool returns WCET information that ranges from \emph{exactly} the WCET of real hardware to 148\% of the new version's measured WCET.
  With the same benchmarks, the unmodified \ait estimates are 112\% and 149\% of the actual executions; therefore, even when \deltatool is pessimistic, an industry-strength tool such as \ait cannot do better.
  Crucially, we also show that \reta decreases aiT's analysis time by 45\% and its memory consumption by 8.9\%, whereas removing \reta from \deltatool, effectively rendering it a regular timing analysis tool, increases its analysis time by 27\%. 

\end{abstract}


\begin{IEEEkeywords}
Software updates, timing analysis, embedded systems, software verification.
\end{IEEEkeywords}

\section{Introduction}

In embedded applications, timing properties are crucial to ensure correctness.
Application requirements dictate the timing properties, yet applications typically execute atop hardware with limited resources, which impacts the ability to meet real-time deadlines.
Missing deadlines may compromise the application's safety properties.

Consider the analysis of worst-case execution time (WCET)~\cite{li_performance_1997}, often carried out based on a model of the program. 
The model may include paths in the code that are actually executed during application runs, and paths that are infeasible in the actual program but considered feasible in the model.
As a result, existing techniques and corresponding tools return \emph{over-approximations} of the actual WCET.
The over-approximation is useful in many respects anyways.
For example, it can be used to determine schedulability since a system that is schedulable with the over-approximated WCET is schedulable also with the actual WCET.

\fakepar{Problem} As codebases undergo evolution and maintenance, ensuring that timing properties are retained after code updates is key to enabling robust development processes~\cite{5591311}.
Software updates usually modify a small proportion of the code.
For example, 75.2\% of commits to the \texttt{gcc} compiler, which has millions of lines of code, change less than 47 lines~\cite{alali_whats_2008}.

Existing analysis techniques are oblivious to evolution and maintenance of existing codebases.
Their application to software updates is the same as with the original program: the code must be analyzed in its entirety, which is computationally expensive and labor intensive.
The latter issue is acute: developers are to provide essential information about the code, such as loop bounds, that may be germane to code sections outside of their expertise, such as hardware drivers.
The effort required is marked even if the effects of the modification on the timing behavior are small or even nonexistent.


\fakepar{Differential timing analysis}
Instead of re-analyzing the entire program to provide absolute timing information, we isolate the effect of updates on the software's timing behavior and only provide \emph{differential timing information}.
We argue that this is sufficient in many settings.
For example, most updates apply urgent bug and security fixes rather than global performance optimizations~\cite{mcilroy_fresh_2016}.
It is sufficient for these cases to provide a guarantee that performance is no worse, or is worse by a small enough amount, than the original program.
Differential timing analysis can provide a guarantee that the new version's performance meets a specific performance target relative to the old version, or demonstrate that this is not the case.
To that end, differential timing analysis is computationally less expensive and does not require as much developer effort as with existing techniques.

Thinking that performing differential timing analysis merely amounts to identifying the lines of codes that change from one version to another is in a \emph{way naive}.
A small code change may, in principle, bear devastating effects on the overall timing behavior.
For example, say an update modifies the value of an integer variable in an assignment.
No other code is modified, hence the net execution time of the changed instructions remains the same.
The variable is used elsewhere in the code to determine the number of iterations in several nested \texttt{for} loops.
The overall increase in execution time is \emph{multiplicative} in the number of nested loops even though the instructions inside the loops have not changed.

\fakepar{\reta}
We develop a differential timing analysis technique called \reta (\underline{Re}lative \underline{T}iming \underline{A}nalysis), described in \secref{sec:reta}.
\reta incorporates a slicing procedure that generates a reduced program with only the lines of code that impact the timing behavior.
In the example above, \reta would identify both (and only) the change in the assignment and the nested loops.
Only these lines of code must be analyzed; therefore, the computational requirements and developer effort are greatly reduced.
\reta further categorizes each statement in the sliced program according to how it impacts the change in execution time across program versions.
The statement's contribution to the change in execution time differs based on the categories.

\reta returns timing information relative to the execution time of the original version of the code.
An example of \reta's output is that the new version is between X and Y time units faster than the old version.
The timing information derived from the original version of the software may  be combined with \reta's output to obtain global timing information.

\fakepar{Practice}
\deltatool (\underline{D}isass\underline{e}mbly \underline{L}evel \underline{T}iming \underline{A}pproximation) is a proof-of-concept semi-automatic implementation of \reta, described in \secref{sec:delta}.
\deltatool disassembles the program binaries and uses a regular diff tool to find differences at the level of assembly code.
Then, the slicing procedure is applied to identify the relevant code segments. 
\deltatool targets microcontroller units (MCUs) of the Cortex M* series, which are often deployed in embedded systems such as nanosatellites and mobile robots.
The output of the slicing procedure is analyzed based on the execution times of individual instructions obtained from the ARM M* reference documentation~\cite{m4_reference_manual_2020}.

We also adapt a subset of \reta to aiT~\cite{ait}, an industry-strength WCET analysis tool, to enable it to analyze updates differentially provided they meet specific conditions.
We call this \aitreta.
As the implementation is only partial, these conditions are significantly more restrictive than \deltatool's.

\fakepar{Results} We consider staple embedded benchmarks and actual updates to the respective codebases, taken from their official code repositories.
We compare the outcome of the analysis obtained with \deltatool with those of \aitreta, the unmodified aiT, and real executions on a NUCLEO-L432KC development board equipped with an STM32L432KC MCU~\cite{nucleo-l432kc}.

Our results, reported in \secref{sec:evaluation}, show that, for example, WCET information returned by \deltatool ranges from \emph{exactly} the WCET observed on real hardware to 148\% of that.
In the same benchmark, \ait measures a  12\% and 149\% higher WCET; \deltatool is thus either more accurate or similarly pessimistic than \ait.
On the other hand, the (partial) implementation of \reta in \ait reduces the analysis time and memory consumption by 45\% and 8.9\%, respectively.

Before moving on to the technical matter, \secref{sec:background} presents a brief survey of related efforts and necessary concepts to understand the rest of the paper.


\section{Background and Related Work} \label{sec:background}


Our work belongs to the field of static deterministic timing analysis~\cite{abella_comparison_2014}.
These approaches are pessimistic by nature.
As mentioned, when computing WCET, the lack of run-time information may guide the analysis to consider certain states as reachable, when they are not.
Program models are combined with device models representing the target platforms.
These models are also generally pessimistic to maintain safety, for example, when representing the behavior of caches and speculative executions.
These features produce estimates that may be significantly worse than the actual values~\cite{mezzetti_randomized_2015}.


Differential timing analysis and \reta intersect the existing literature in many ways.
We provide next a brief account of the works we deem close to ours.


\subsection{Update Analysis}

The idea of analyzing code changes per se, rather than analyzing entire programs, is applied to various problems.

Differential symbolic execution is a technique that utilizes differences in a program to characterize specific behavioral differences or verify behavioral equivalence ~\cite{person_differential_2008}.
The motivations are similar to our work, in that regular symbolic execution does not consider that updates typically modify a small portion of the code~\cite{person_differential_2008}.
In contrast to our work, differential symbolic execution does not verify timing properties, but rather functional equivalence, meaning that the input-output relations are retained across both versions

A related technique is regression verification, which verifies program equivalence as an alternative to verifying functional correctness~\cite{godlin_regression_2009}.
The rationale is that the verifier no longer needs to know the properties that the program must meet, but rather only checks that the updated program meets the same properties as the original one.
Program equivalence again holds when the input-output relations are identical between the two versions.
Every sorting algorithm is equivalent according to this definition.
However, different sorting algorithms (and their implementations) are not equivalent from a timing perspective, which differential timing analysis can instead verify.




\subsection{Implicit Path Enumeration}
\label{implicit_path_enumeration_technique}

Static deterministic timing analysis is normally performed using implicit path enumeration~\cite{li_performance_1997}.

First, micro-architecture analysis builds a control flow graph (CFG) of the program~\cite{raymond_general_2014}.
Nodes represent basic blocks; edges represent  jumps from one basic block to another.
For each node, a WCET is computed.
Next, data flow analysis uses the CFG to determine feasible paths and loop bounds~\cite{raymond_general_2014}.
In doing so, it generates program functionality constraints that define bounds on the values of variables using both variable definitions and branching conditions.
An optimization problem is then solved to maximize the total WCET subject to program functionality constraints~\cite{raymond_general_2014}.

\figref{fig:example_data_flow} shows an example representing a \code{while} loop.
Node~1 ends with a conditional jump: if variable \code{x} is less than 5 then the execution jumps to node 2, otherwise it continues to node 3.
Node~2 ends with an unconditional jump to node 1, represented by an edge from node 2 to node 1.
The \code{while} loop repeats until the variable \code{x} is 5 or larger.

Data-flow analysis obtains loop bounds by analyzing the possible values of a variable, with knowledge of the code inside each node.
Let \code{x} be a positive integer or zero at the start.
Let the code in node 2 increment \code{x} by one.
The analysis places a constraint on the maximum number of times node 1 executes, which in this case is six times, and would occur when \code{x} starts with 0. 
In \figref{fig:example_data_flow}, this occurs when node 1 executes six times, node 2 executes five times, and node 3 executes once.

\begin{figure}[tb]
    \centering
    \includegraphics[width=0.25\textwidth]{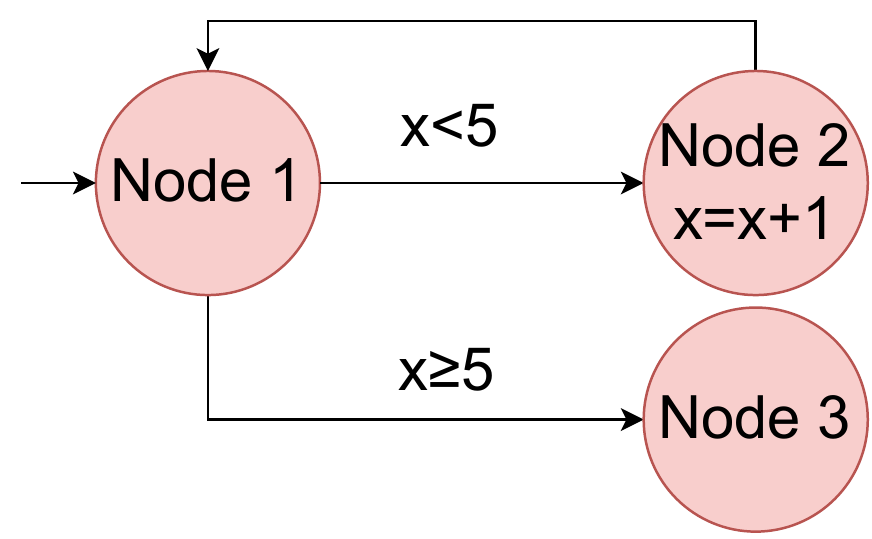}
    \caption{The CFG of a \code{while} loop.}
    \label{fig:example_data_flow}
    \vspace{-4mm}
\end{figure}

\subsection{Abstract Interpretation} \label{abstract_interpration}

Abstract interpretation is a technique for safely approximating the behavior of programs~\cite{cousot_abstract_1977}.
It obtains abstract semantics by performing an abstraction of the concrete values. For example, interval abstraction represents a variable's values with an interval defined between minimum and maximum values.
To apply this abstraction safely, the abstract interval must contain all values the concrete variable may be assigned.
For example, if the value of a concrete variable \code{x} is in a set \{1, 7, 42\}, then a safe interval abstraction is the integer range [1,42].
The abstraction is less precise than the concrete semantics,  in that it includes values that the concrete variable cannot take, such as 3 or 11 in this example. 

Abstract interpretation is used when analyzing the program with concrete values is not feasible.
Interval analysis, for example, allows one to use interval arithmetic to establish constraints on the values of different variables without the need to analyze every possible concrete value.
Additionally, computing a minimum and maximum value for a variable is usually less expensive than computing all possible values.


Abstract interpretation is used in WCET analysis to compute the values a variable may have, which aids in finding loop bounds~\cite{ermedahl_loop_2007} and in determining the reachability of code.

\subsection{Program Slicing}

Program slicing is a technique to reduce a program into only the instructions that have an effect on the semantics of interest~\cite{harman_overview_2001}.
For example, if we want to determine a variable's possible values, then the corresponding program slice only includes instructions that influence the variable's values in some way.
The variable at hand is known as the slicing criterion.
The dependency is transitive, meaning that if the slicing criterion is influenced by a variable \code{x}, and variable \code{x} is influenced by variable \code{y}, then the slicing criterion is influenced by variable \code{y}.
This approach is known as backward slicing~\cite{silva_vocabulary_2012} because the analysis moves backwards to find the statements that influence the value of our slicing criterion.

The opposite approach is forward slicing~\cite{silva_vocabulary_2012}.
Forward slicing moves in the forward direction to find the statements whose variables depend on the value of the slicing criterion.
This dependency is also transitive.
The forward slice can be used to determine how a change in the value of the slicing criterion propagates through the rest of the program.


\section{Traditional timing analysis example}
We attempt to conduct an analysis of Hackflight~\cite{levy_robustness_2020}, an open-source drone autopilot, using aiT. We note the challenges associated with using traditional WCET analysis for embedded software.

Hackflight is an open-source drone autopilot. A drone autopilot controls a drone by obtaining data from sensors, such as the current orientation, and uses it along with the pilot's inputs to move the drone in a particular orientation and speed. A drone autopilot typically does so using a PID (proportional-integral-derivative) controller which attempts to reduce the error between the pilots inputs (intended orientation and speed) and the drone's actual orientation and speed.

Hackflight has about 4000 lines of code, which makes it significantly smaller than other autopilots such as Ardupilot~\cite{tridgell_2022}. A typical Ardupilot binary is on the order of megabytes, whereas a Hackflight binary is on the order of tens of kilobytes. 

We choose to analyze Hackflight because of its small size, which informs us on how existing WCET techniques scale to real-world embedded software.

We conduct a WCET analysis of Hackflight. We specifically analyze Hackflight's main control loop. A single iteration of Hackflight's main loop obtains pilot inputs by reading data from a radio receiver, obtains data from sensors, and runs a PID control algorithm to find appropriate motor power magnitudes to achieve the orientation specified by the pilot.

We find that conducting a WCET analysis of an embedded program is difficult even with sophisticated and mature tools. We conduct the WCET analysis using aiT. 

We find that analyzing the entire program requires us to provide a significant amount of necessary information. Some of the information we provide include:
\begin{itemize}
    \item Loop bounds (which cannot always be determined automatically).
    \item Branch instructions where the target is an address in a register (most notably used by virtual functions in C++ and function pointers).
    \item Access timing for different memory regions.
    \item Measured/Documented execution times for input/output from sensors.
\end{itemize}

Some information may be extremely difficult for developers to provide especially information about external libraries, such as hardware drivers. Some of the loop bounds we must provide to aiT are bounds for loops used in hardware libraries such as timers, I\textsuperscript{2}C, and C++ functions such as \textit{\_malloc\_r} and \textit{memcpy}. 



As we demonstrate in \secref{sec:reta}, we can identify the timing impact that a code change has on other sections, and use this to compute the change in execution time as a result of an update. Doing so requires significantly less computational and user effort than traditional timing analysis, since an update usually only affects a small proportion of the code.

\section{Relative Timing Analysis} \label{sec:reta}

\reta (\underline{Re}lative \underline{T}iming \underline{A}nalysis) is our take at differential timing analysis.
\reta returns the change in execution time from the original to an updated version of a software, given the same program inputs.
If this figure is a positive value, then the updated version is slower than the original version; if it is negative, then the updated version is faster.
We explain next the requirements that must be met for \reta to be applicable, as well as the analysis procedures at its core.

\subsection{Requirements} \label{reta_requirements}

\reta requirements are largely the same as most timing analysis techniques and tools.

\reta must operate at the level closest to actual program execution, that is, with assembly code.
Working at source code level would mislead the estimates due to compiler optimizations~\cite{li_traceability_2014}.
To apply the slicing procedure we describe next, branches or jumps must have statically-known targets or the user must provide the target.
We also require the line-by-line differences between the two program versions.
Equivalent lines must also be provided, where two equivalent lines have a one to one mapping and are semantically equivalent.
Existing differencing algorithms~\cite{hunt_algorithm_1976} can determine this information.

The execution times of each instruction must be known, for example, in terms of clock cycles, including instructions accessing peripherals.
These may be obtained from data sheets or measured on the hardware.
If exact timing information is not available, at least a lower- and upper-bound must be available.

We target single-core platforms, which are commonly employed in embedded systems such as closed-loop control equipments, mobile robots, and nanosatellites~\cite{7321580,7119042}.
The interleaving between different cores in multicore systems would make it difficult to reason about the effect of specific instructions, since the time of instruction execution must also be considered.
We also consider instructions to execute sequentially, and therefore do not incorporate out-of-order execution, which may result in different execution times than a sequential model. 

\begin{figure}
    \centering
    \includegraphics[width=0.49\textwidth]{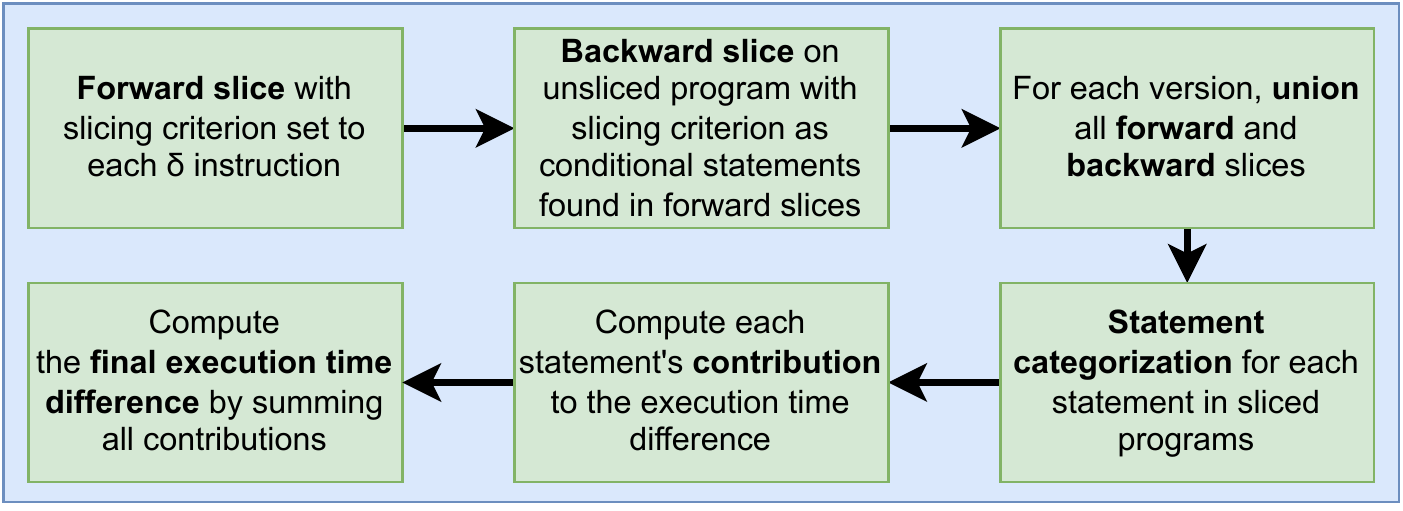}
    \caption{Timing analysis of software updates with \reta.}
    \label{fig:reta_procedure_figure}
    \vspace{-1em}
\end{figure}

\subsection{Analysis} \label{reta_procedure}

We provide an overview of the analysis procedure, with \figref{fig:reta_procedure_figure} serving as a road-map.
We illustrate a concrete example in \secref{sec:delta} when this procedure is concretely applied to example code using our \deltatool implementation.

The procedure takes as input the code differences determined by a regular differencing algorithm.
We call every such instruction as \deltainstruction.
We slice both program versions \emph{forward} from every \deltainstruction to determine the later instructions that are possibly affected.
For each conditional instruction in the forward slice, we slice the corresponding program version \emph{backward} to obtain complete information on their truth values.
The combination of forward and backward slices for a given version of the code, over all \deltainstructions, includes instructions that either \emph{i)} are different between program versions,  \emph{ii)} whose execution is possibly impacted by  code changes, or \emph{iii)} concur to determine execution paths that might change because of code changes.



The two program slices, one for the original program and one for the updated program, determine the instructions of interest to compute the time difference.
Different types of instructions contribute differently to estimating this difference.
We categorize each instruction in a slice as:
\begin{enumerate}
    \item CatA: sequential execution, that is, instructions that do not split the program flow. 
    \item CatB: branching points, that is, instructions that do split the program flow; in a CFG, these instructions appear last in a node that has two or more outgoing edges:
    \begin{enumerate}
        \item CatB\textsubscript{a} are conditional instructions that do not cause a natural loop, that is, there is no path in the CFG that returns to the same instruction.
        \item CatB\textsubscript{b} are conditional instructions that cause a natural loop, represented in a CFG as a circular path that eventually returns to the same instruction.
    \end{enumerate}
\end{enumerate}

Each category requires a different analysis, as exemplified in \secref{sec:delta}, that determines the instruction's individual contribution to the time difference, which may be negative or positive.
The final execution time difference is the (arithmetic) sum of the individual contributions. 
\nottoggle{extended}{
    An extended version of \reta's procedure is in an accompanying technical report~\cite{yaacoub2023timing}.
}{}

\iftoggle{extended}{
    \begin{figure*}[t]
        \centering
        \includegraphics[width=\textwidth]{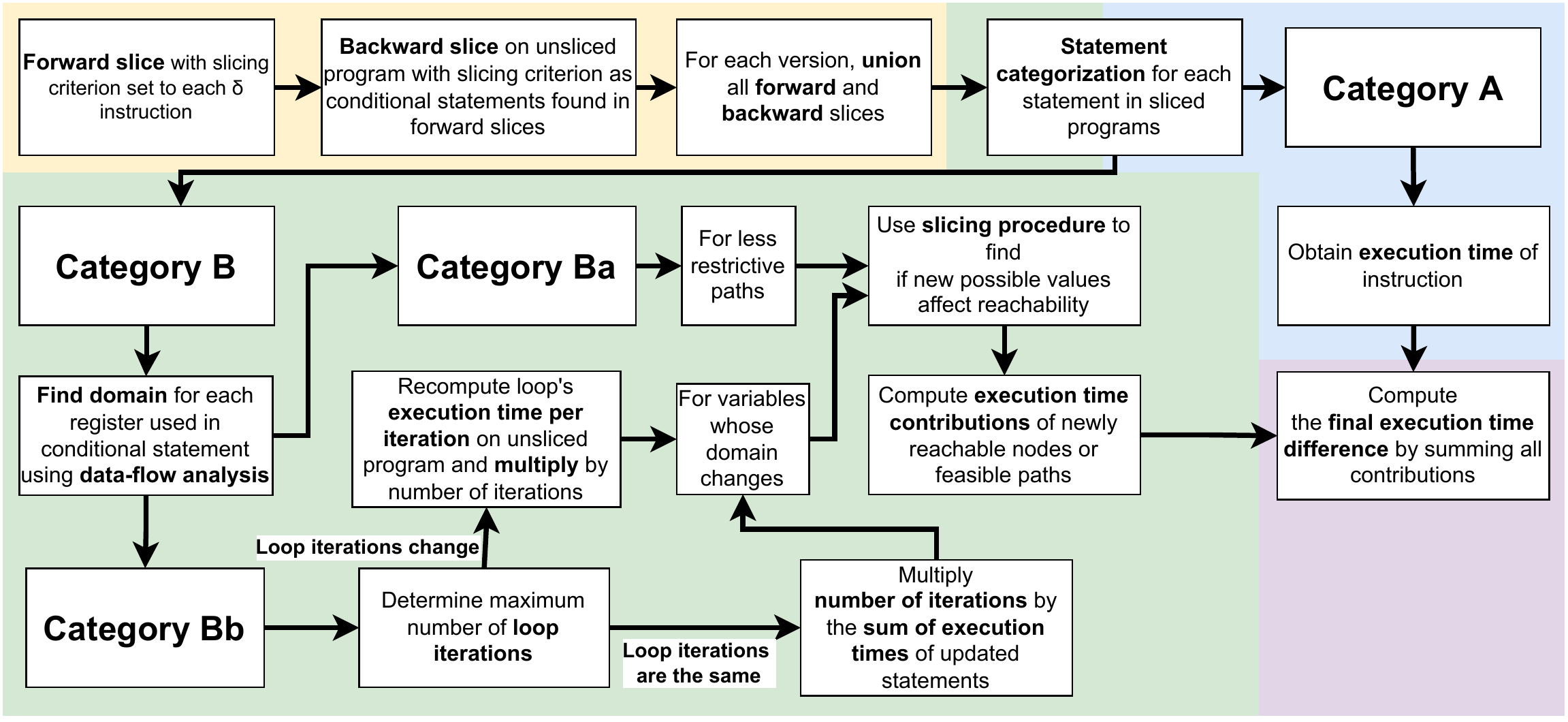}
        \caption{Extended \reta procedure}
        \label{extended_reta_procedure_figure}
    \end{figure*}

    \figref{extended_reta_procedure_figure} demonstrates a more detailed procedure for \reta that highlights the steps that must be taken for specific instruction categories as well. The figure is color coded to highlight key steps. The slicing procedure is in yellow. The statement categorization procedure is in blue, and green. Blue corresponds to Category A and green corresponds to Categories B\textsubscript{a} and B\textsubscript{b}. The final step of the procedure is purple.
}{}


\section{\deltatool} \label{sec:delta}

\deltatool (\underline{D}isass\underline{e}mbly \underline{L}evel \underline{T}iming \underline{A}pproximation) is an implementation of \reta that targets ARM Cortex M4F microcontrollers.
The M4F is the lowest-power microcontroller of the Cortex M* family that is also equipped with a hardware FPU.
This makes it applicable to embedded systems that are both energy constrained but also require computing power, as in low-level robot controllers~\cite{bregu_reactive_2016}.

We provide here essential information on the implementation of \deltatool.
Next, we walk the reader through a complete example of differential timing analysis.
\iftoggle{extended}{ We provide the reader with an additional example that covers a special case.}{}
\nottoggle{extended}{
More complete information, including an additional more sophisticated example of differential timing analysis, are included in an accompanying technical report~\cite{yaacoub2023timing}.}{} 
We conclude with a discussion of the factors affecting the accuracy of \deltatool.

\subsection{Implementation Highlights}

The analysis requires the ELF file (Executable and Linkable Format) of the original and the updated software.
We convert both to assembly code for the target architecture using the \code{objdump} tool included within the ARM GNU toolchain~\cite{arm_gnu_toolchain} for ARM Cortex M* microcontrollers.
The assembly code of the original program is compared to that of the updated program using the differencing tool included within Visual Studio Code to identify code differences and equivalent lines. 

We conduct the slicing procedure using a mixture of programmer-provided slicing information and the slicing tool included within Ghidra, a reverse engineering tool \cite{ghidra}.
Ghidra decompiles the assembly code to C code on a function level.
As a result, program slicing within Ghidra only slices inside a function (intraprocedural slicing), rather than throughout the entire codebase (interprocedural slicing). 

Once the slices are obtained, loop bounds and variable domains are determined using data-flow analysis if necessary, or based on programmer-provided knowledge.
Backward slicing may not be necessary if loop bounds are programmer-provided.
The execution times of individual instructions are obtained from the ARM M4 reference manual~\cite{m4_reference_manual_2020}.

Programmers participate in the slicing procedure.
The automated part of the slicing procedure uses Ghidra to slice on a decompiled version of the program binary.
However, since Ghidra does not slice across different decompiled functions, the programmer must provide where in the binary the slicing procedure should continue.
Programmers also need to perform the dataflow analysis required to compute loop bounds and determine reachability.
Alternatively, they can provide loop bounds based on application-level knowledge for loops found in the forward slice.
\begin{figure*} [tb]
    \centering
    \begin{subfigure}[b]{0.23\textwidth}
      \centering
        \includegraphics[width=0.9\textwidth]{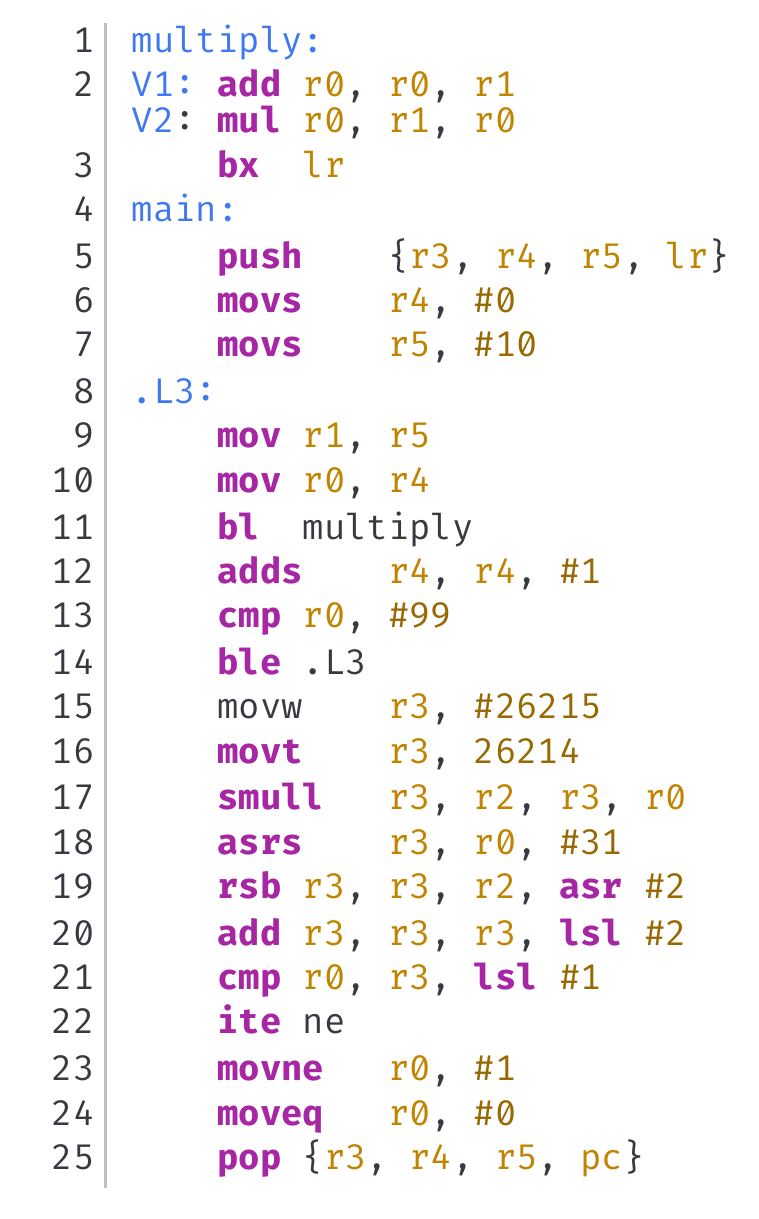} 
        \caption{An update fixes an \\ incorrect multiplication \\ routine. \\}
        \label{code_before_slicing}
    \end{subfigure}
    \begin{subfigure}[b]{0.23\textwidth}
        \centering
        \includegraphics[width=0.9\textwidth]{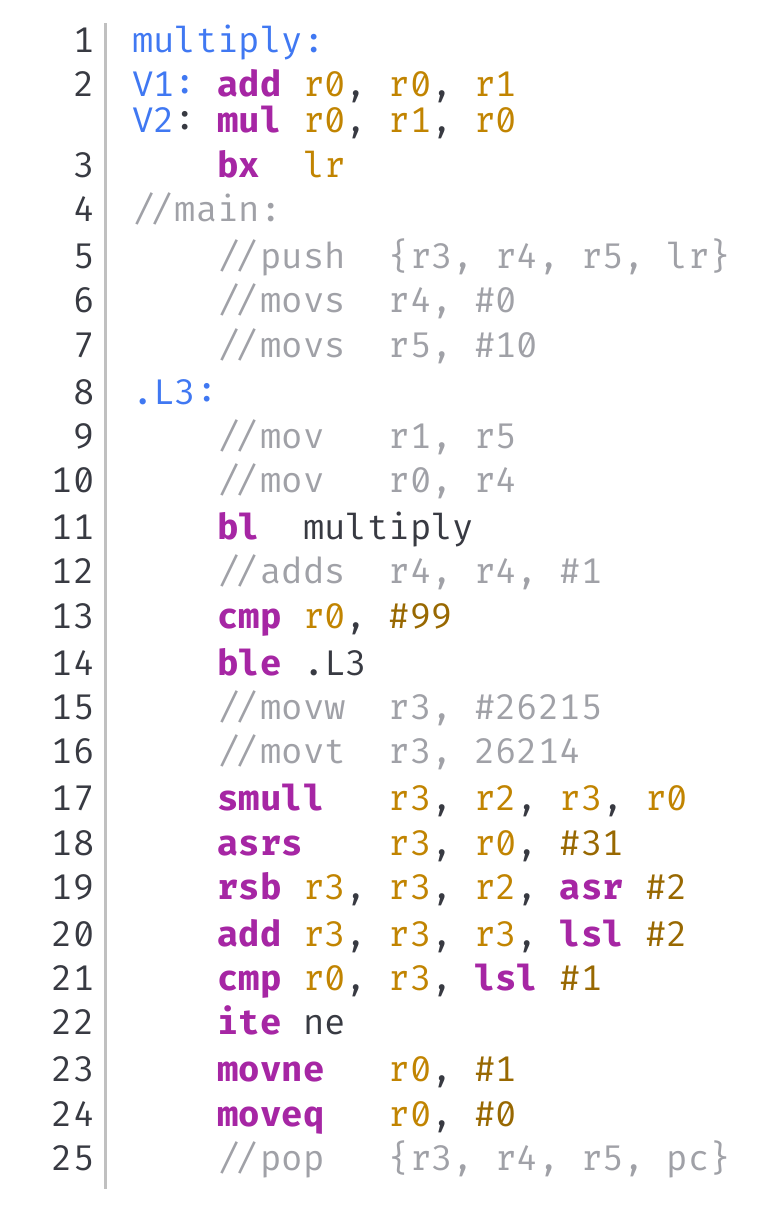} 
        \caption{Forward slice of the \\ code in \figref{code_before_slicing}; the slicing \\ criterion is the update \\ in line 2.}
        \label{code_after_forward_slice}
    \end{subfigure}
    \begin{subfigure}[b]{0.23\textwidth}
        \includegraphics[width=0.9\textwidth]{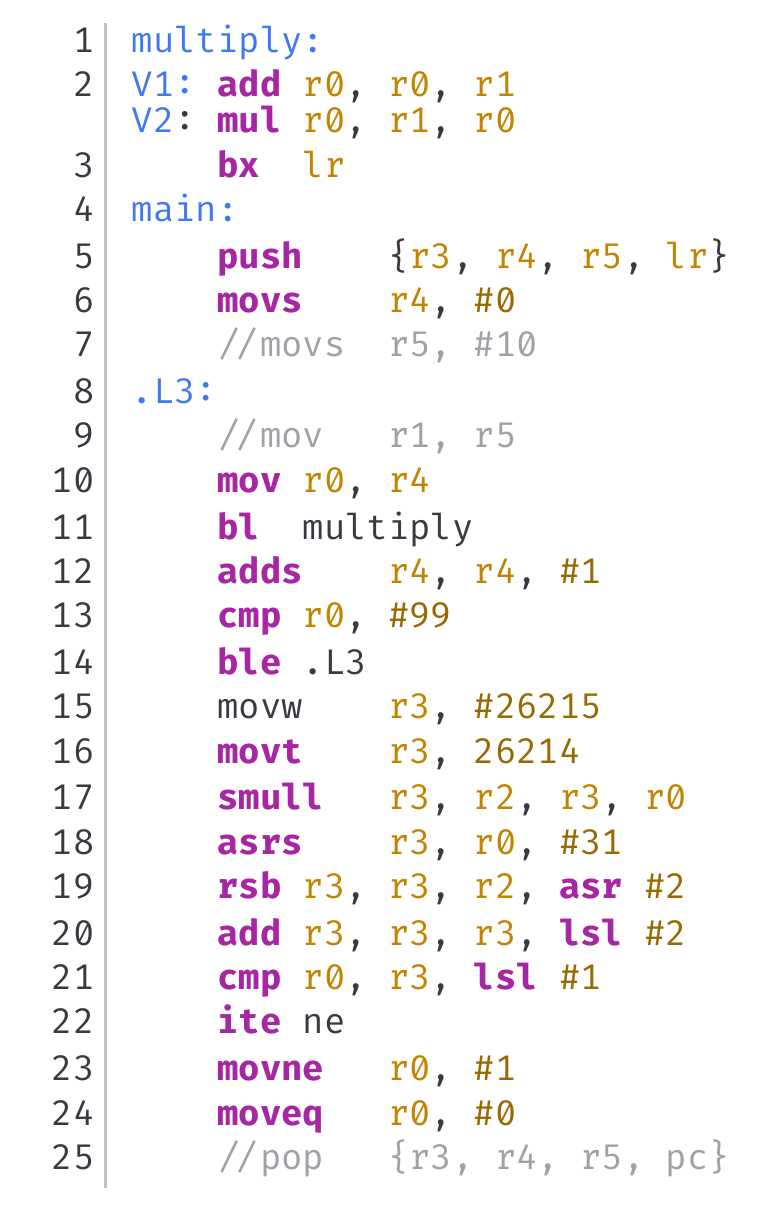} 
        \caption{Backward slice of \\ \figref{code_before_slicing} driven by conditional \\ instructions of the \\ forward slice.}
        \label{code_after_backward_slice}
    \end{subfigure}
    \begin{subfigure}[b]{0.23\textwidth}
        \includegraphics[width=0.9\textwidth]{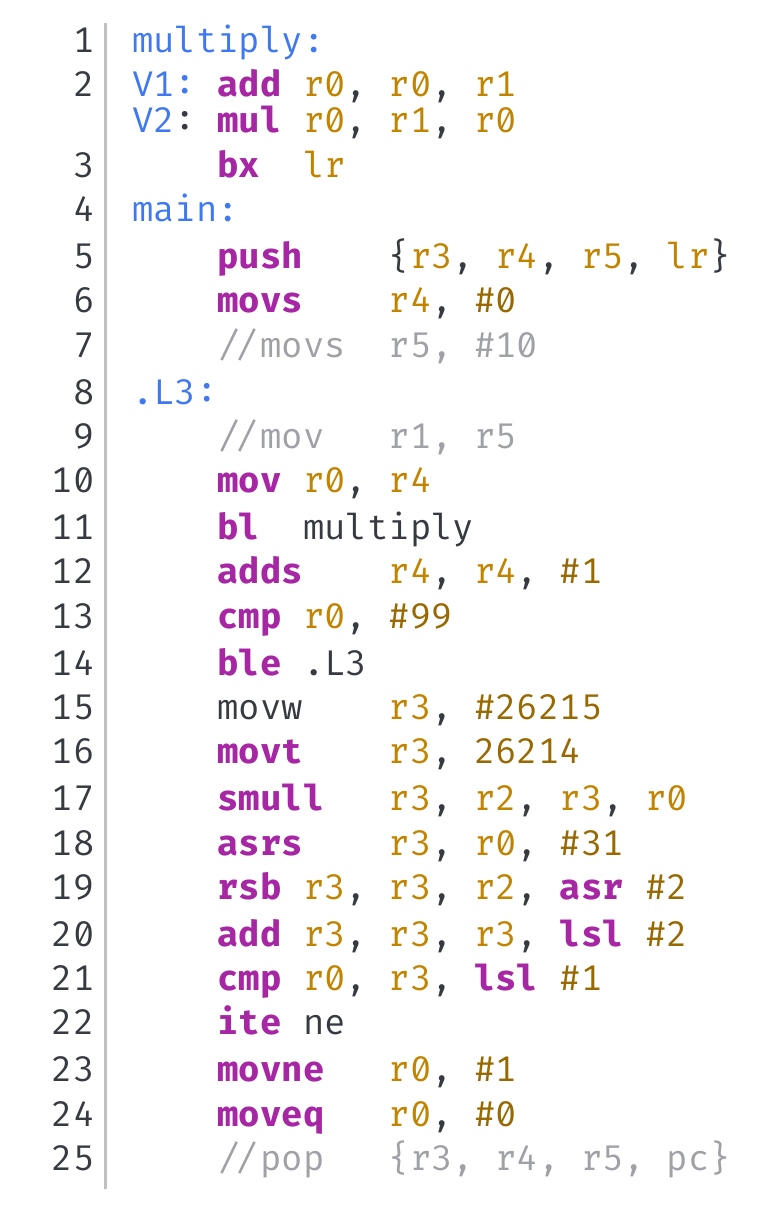} 
        \caption{Combining forward and backward slices in \figref{code_after_forward_slice} and \figref{code_after_backward_slice} for categorization. \\}
        \label{code_after_slicing}
    \end{subfigure}
\caption{Assembly code at different stages of the slicing procedure. Commented lines do not belong to a specific slice.}
\label{fig:assembly_codes}
\end{figure*}

\subsection{Example Analysis}
\label{applying_delta}

Consider the assembly code in \figref{code_before_slicing}.
We are to find the execution time difference caused by fixing an incorrect multiplication routine, which used the \code{add} instruction instead of the \code{mul} on line~2.
The update fixes the bug.

The differencing tool indicates that the only difference between the two programs is in line 2.

\fakepar{Slicing} We apply the slicing procedure explained in \secref{sec:reta} to \figref{code_before_slicing}.
First we compute the forward slice with slicing criterion set to line~2.
The resulting program slice, shown in \figref{code_after_forward_slice}, includes all later instructions impacted, directly or indirectly, by a change in register \code{r0}.
For example, it includes the computation of a new value for \code{r3}, as in line 18.
The slice also includes conditional instructions on lines 13 and 14, which means we need the slice to include the taken and not taken paths. 
Lines 21 and 22 are also conditional instructions and a similar reasoning applies there. 

Next, we perform a backward slice of either program version with the slicing criterions set to the conditional instructions possibly included in the forward slice. We set the slicing criterions to lines 13, 14, 21, and 22. Lines 21 and~22 determine which of lines 23 and 24 are executed. 
The resulting backward slice is in \figref{code_after_backward_slice}.
We find that \code{r1} and \code{r5} do not determine the outcome of either conditional instruction, and \code{r3} only affects the conditional instruction at line 23 and 24, starting from the values it is assigned at lines 15 and 16.

The final sliced program, shown in \figref{code_after_slicing}, combines the forward and backward slices and is the input to the categorization we discuss next.
For brevity, we do not discuss the categories for all instructions in \figref{code_after_slicing} but instead concentrate on illustrative examples for given categories.

\fakepar{CatA} The first instruction of the main label is a push instruction (line 5).
This instruction does not split the control flow of the program and therefore, its effect on timing is solely due to its own execution time: it belongs to CatA as described in \secref{sec:reta}.
Since the instruction is the same in both versions and on equivalent lines, it has no impact.

A more interesting example is the only instruction the programmer changes in the updated program, found in line 2.
This instruction changes from an \code{add} to a \code{mul} instruction.
Since each of those two exists in one version but not the other, they impact the execution time difference if they have different execution times. Instructions that only appear in the original version bear a negative contribution, whereas instructions that only appear in the updated version bear a positive contribution.
In ARM M4 microcontrollers, both \code{add} and \code{mul} take one cycle. Their relative contributions cancel each other out. 

\fakepar{CatB\textsubscript{a}} Consider the conditional instructions on lines 21 and~22.
These instructions split the control flow depending on the value of the register(s) being compared.
Therefore, the update may introduce new behaviors here.
We can determine whether this is the case through data-flow analysis.

For illustration's sake, we consider the source code equivalent to \figref{code_before_slicing}, shown in \figref{source_code_before_slicing}, to discuss the use of data-flow analysis in this example.
The conditional instruction in line~13 of \figref{source_code_before_slicing} evaluates to false when the result of the multiply function is a multiple of 10 and true otherwise.
We do not yet know when the loop terminates, so we can assume it may terminate at any non-negative value of \code{i} for now.
Using data-flow analysis, we find that the update makes the \code{true} case impossible, since the output of the \code{multiply} function must be a multiple of 10.

We infer that the update makes the \code{true} condition of the conditional instruction more restrictive, that is, the set of possible values for variable \code{result} that return \code{true} for the conditional instruction in the updated version is smaller than the same set in the original version.
We can then exclude the \code{true} branch from the analysis since we know that no new possible values of \code{result} exist and therefore, no new behavior is introduced.
This property holds provided that the domains of all variables in the updated version are more restrictive or equal to the original version's.

\begin{figure}
    \centering
    \includegraphics[width=0.3\textwidth]{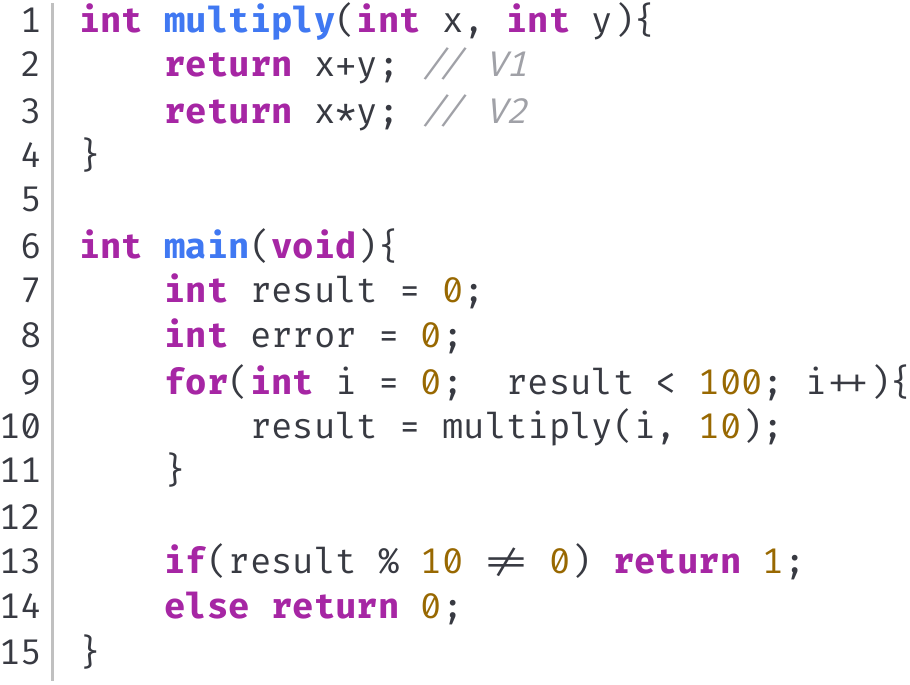}
    \caption{Source code for \figref{code_before_slicing}.}
    \label{source_code_before_slicing}
\end{figure}

Conversely, the update made the \code{false} condition of the conditional instruction strictly less restrictive.
This is because the domain of \code{result} in the updated version is strictly larger, that is, it includes additional values not present in the original version's domain.
For example, if the loop terminates when \code{i} is 5, the value of \code{result} differs between both versions. The updated version evaluates to false while the original version does not.
Therefore, there are values that evaluate to \code{false} in the new version, but not in the original one.

As a result, there may exist new behaviors.
To understand that, we only need to analyze the values present in the updated version's variable domain but not the old version's.
Values present in both do not result in new behaviors.
Only analyzing new values significantly decreases the required analysis when compared with traditional timing analysis, which always needs to analyze all possible values.
Since both paths (lines 23 and 24) only have a single instruction with identical execution time, there is no contribution to timing difference.

\fakepar{CatB\textsubscript{b}} Consider the branching instruction on line 14.
This is part of a loop, since the code in lines 8-14 repeats while \code{r0} is less than or equal to 99.
Therefore, a change in how \code{multiply} computes its output may affect the number of loop iterations, which changes the execution time.
We investigate this with data-flow analysis, again using source code for easier illustration.
With variable \code{i} in \figref{source_code_before_slicing} initialized to 0, in the original version the loop terminates after 90 iterations.
Therefore, the maximum number of loop iterations is 90. In the updated version, the loop terminates once \code{i} reaches 10, which happens after 10 iterations. 

We compute the execution time of a single loop iteration from the assembly code of \figref{code_before_slicing}.
Each iteration requires 17 cycles in both versions of the software.
In the original version, the loop executes 90 times.
Its contribution to differential timing analysis is negative and amounts to the product of each iteration's execution time and the number of executions, leading to a (negative) contribution of 1530 cycles.
In the updated version, the loop executes 10 times.
Its contribution to differential timing analysis is now positive and amounts to 170 cycles total, as per the previous reasoning.
The execution time difference is then -1360 cycles, that is, the update improved the loop's performance by 1360 cycles.

Note how the domain of variable \code{i} in \figref{source_code_before_slicing} changes from [0,90] to [0,10].
Every value of \code{i} that is possible in the updated version is possible in the original version, and therefore no new behavior is introduced.
If the domain of the updated version had been larger than the original version, we would also check whether \code{i} is involved in conditional instructions after the loop terminates.
The slicing procedure would identify these cases.

\fakepar{Summing up} Through the analysis of the assembly code of \figref{code_before_slicing}, we discover that \emph{i)} the lines the programmer changes in the updated version bear no contribution to the timing difference, \emph{ii)} the conditional instructions on lines 21-24 retain the original execution times also after the update, \emph{iii)} the loop in lines 8-14 changes the number of iterations, which reduces the loop execution times by 1360 cycles, and \emph{iv)} all other instructions have identical timing contribution across both versions.
Then we conclude that the execution time difference is -1360 cycles, which means the update made the program as a whole 1360 cycles faster. This is a reasonable result because the number of loop iterations has decreased by a factor of nine and each iteration is fairly small at only six instructions.

\iftoggle{extended}{
    \subsection{Addition of new conditional statements} \label{addition_new_conditional_statements}

    \begin{figure}
        \centering
        \begin{minipage}{0.24\textwidth}
            \centering
            \includegraphics[width=0.9\textwidth]{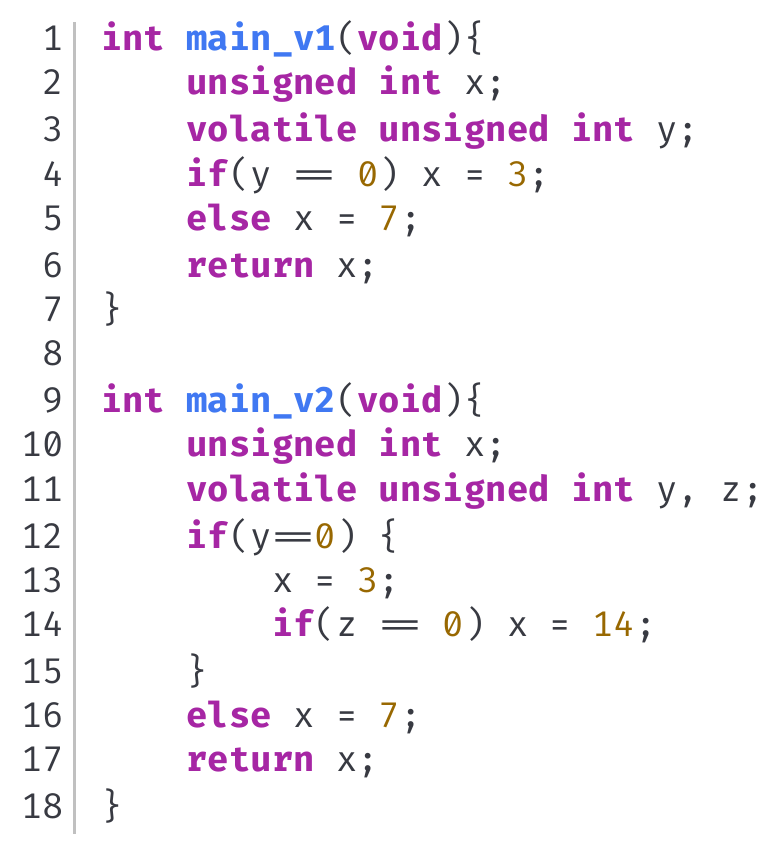} 
            \caption{Category B\textsubscript{a} example where the CFG graph is changed through the addition of a new node and edges}
            \label{category_b_a_2_code}
        \end{minipage}\hfill
        \begin{minipage}{0.24\textwidth}
            \centering
            \includegraphics[width=0.9\textwidth]{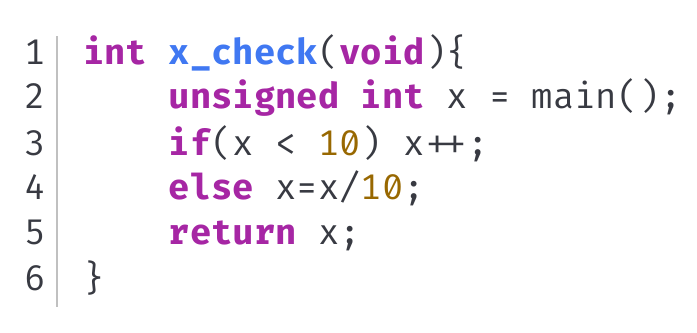} 
            \caption{A function that uses the output of \figref{category_b_a_2_code} in an if statement}
            \label{category_b_a_2_additional_code}
        \end{minipage}
    \end{figure}

    \figref{category_b_a_2_code} demonstrates a more complex example where the update adds an extra conditional statement, which corresponds to adding an extra node in the CFG, along with any necessary edges. We demonstrate the old CFG in \figref{fig:Category_Ba_example_CFG} and the new CFG in \figref{fig:Category_Ba_example_CFG_v2}. We highlight the added/removed nodes/edges in red. The update requires the addition of one node and three edges, and the removal of one edge.

    The differences on the assembly code level are the following: version 1 uses two conditional move instructions to set x to either 3 or 7 depending on whether the condition $y==0$ is satisfied or not. Version 2 has a branching instruction that can allow program to take the y!=0 path, which sets x to 7. If the y==0 path is taken, version 2 uses a conditional move instruction to set x to either 3 or 14 depending on whether the condition $z==0$ is satisfied or not. 

    The slicing procedure yields that the changes impact all conditional statements because on the assembly code level, the first control flow split is implemented differently (a conditional move in version 1 and conditional branch in version 2). To compute the execution time difference, we compute the execution time of the added statements, and subtract the execution time of the removed statements. Since we have multiple newly added paths, we determine the difference from the longest path, i.e., the path that results in the largest execution time difference. 

    We also need to check the variable domains using data-flow analysis. For example, in \figref{fig:Category_Ba_example_CFG}, x may be set to 3 or 7, but in \figref{fig:Category_Ba_example_CFG_v2}, x may be set to 3, 7 or 14. Therefore, there is one additional possible value. We must analyze whether setting x to 14 introduces new paths that were not accessible in the old version, and if so, we must analyze their effect on the execution time. This depends on where the value of x and hence the return value of the function is used. If we assume this to be the entire program, then no new behavior is introduced. 

    If the return value of the function is used, then that usage will appear in the sliced program. In \figref{category_b_a_2_additional_code}, we demonstrate an example of a second function that uses the return value of \figref{category_b_a_2_code}. It is used in an if statement, and determines where an addition or division is performed. The division is only reachable if the return value is 10 or greater. Since possible values of x before the update were 3 or 7, the division would not be reachable, but since the update introduces an additional possible value of 14, the division is now reachable. Division is slower than addition. Therefore, at specific inputs, there is an execution time difference due to an alternative path being taken. The inputs are when x is set to 14, which is when y and z are set to 0. In the old version, z does not exist and when y is set to 0, x is set to 3 and addition is performed. In the new version, division is performed. The extra execution time of division must be computed to find its contribution towards the execution time difference specifically for the case when y and z are set to 0.

    \begin{figure}
        \centering
        \includegraphics[width=0.5\textwidth]{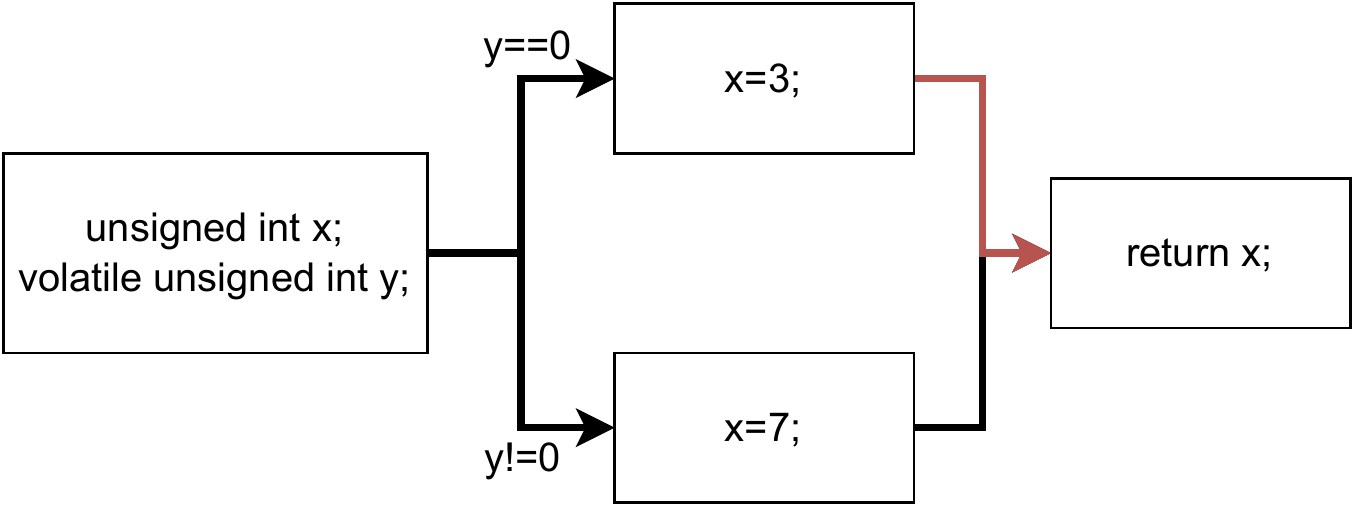}
        \caption{CFG for old version of category B\textsubscript{a} example. The removed edge is highlighted in red.}
        \label{fig:Category_Ba_example_CFG}
    \end{figure}

    \begin{figure}
        \centering
        \includegraphics[width=0.5\textwidth]{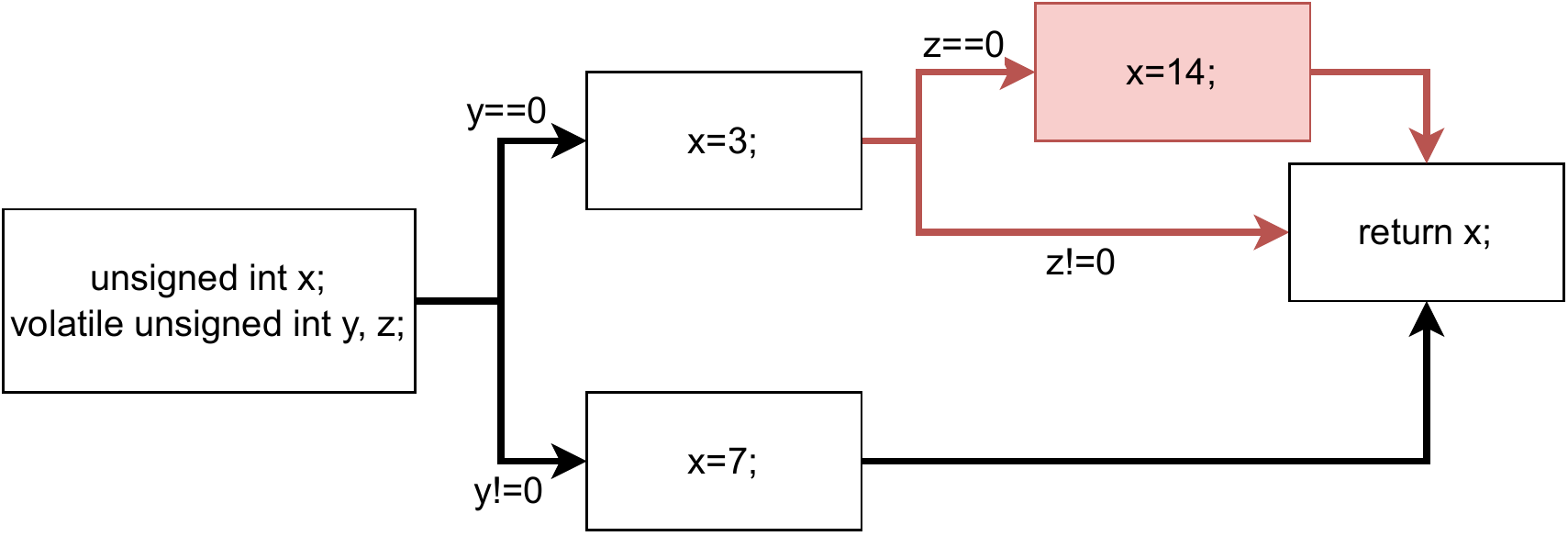}
        \caption{CFG for new version of category B\textsubscript{a} example that adds a node and 3 edges, highlighted in red.}
        \label{fig:Category_Ba_example_CFG_v2}
    \end{figure}
}{}

\subsection{Accuracy} \label{analysis_imprecision}

Three aspects are key when reasoning on the accuracy of the information returned by differential timing analysis in \deltatool.

\fakepar{Variable timings} Many computing cores feature instructions with variable execution times, including the ARM Cortex M4F core we consider~\cite{m4_reference_manual_2020}.
This is not an issue in traditional timing analysis because using the upper  bounds provides a safe over-approximation nonetheless.
For example, by using the upper-bound execution times, the computed WCET is guaranteed to be as large or larger than the actual WCET.

When performing differential timing analysis, however, further care must be taken.
When computing the change in execution time, over-approximating both the speed-up of removing instructions and the slow-down of adding instructions does not necessarily provide a safe over-approximation of the actual execution time difference, but rather something in between that does not suffice to provide robust guarantees on the timing behavior of the updated software.



To address this issue, whenever instructions with variable execution times are involved, differential timing analysis must provide two outcomes.
We determine the \emph{pessimistic} execution time difference using the maximum slow-down, which is an over-approximation, and the minimum speed-up, which is an under-approximation.
Likewise, we determine the \emph{optimistic} execution time difference using the minimum slow-down and the maximum speed-up.


The pessimistic outcome provides a guarantee on the worst-case performance degradation caused by the update.
It can guarantee that a performance target is met, and therefore that the update is safe to deploy from a timing perspective.
It is analogous to WCET in traditional timing analysis.
The optimistic outcome provides the best-case performance improvement due to the update, which is useful during development.
If the optimistic difference is smaller than the target speed-up, then the target is guaranteed to not be met yet, and therefore performance optimizations should continue.
It is analogous to the best-case execution time in traditional timing analysis.
The two measures, taken together, indicate the interval the execution time difference falls within. 

For a fixed execution time difference, the original WCET can be obtained by applying the analysis backwards, that is, by subtracting the execution time difference from the new WCET. However, if the same procedure is applied with optimistic and pessimistic execution time differences, then the original WCET computed is also an interval. Therefore, the original WCET cannot be obtained again, but the original WCET is guaranteed to lie in the computed interval.


\fakepar{Sources of inaccuracy}
\deltatool does not account for caches, pipelines, wait states, and branch prediction. The lack of cache analysis can significantly degrade results for systems with large caches.
Hardware features such as speculative execution are also not taken into account, and may lead to inaccurate results for systems where these play a major role.

Since \deltatool assumes sequential execution, asynchronous events such as interrupts are difficult to analyze. Static timing analysis in general is unsuitable for software with interrupts since they can trigger at any point in the program. \deltatool is still useful with interrupts, however, because \textit{if} the interrupt service routine is modified, \deltatool can identify the corresponding change in execution time, and therefore the impact each interrupt trigger bears on the timing.

\deltatool only uses the ARM M4 reference manual as a source of information on the instruction execution times~\cite{m4_reference_manual_2020}.
For instructions with variable execution times, \deltatool also assumes that instructions that are not in the sliced program have the same execution time across both versions.
Therefore, they cancel each other out.
This likely holds for most instructions but may be violated by instructions that have high spatial or temporal locality to instructions in the sliced program, as the behavior of caches and pipelines may affect their execution times.

These sources of inaccuracy are not a property of \reta, but instead of our specific \deltatool implementation.
Many of the issues are well-researched topics nonetheless.
For example, analysis techniques exist on caches~\cite{mueller_timing_2000} and pipelines~\cite{engblom_processor_2002}. Most of these techniques may be integrated in \deltatool, whereas here we focus on demonstrating the essential ideas behind \reta: the slicing and categorization procedures. 

Due to spatial and temporal locality, changes in the cache behavior are likely to be restricted to instructions that either execute shortly before or after the instructions affected by the update, or instructions that access data regions nearby to those accessed by instructions affected by the update.
Based on this observation, some form of differential analysis can be developed to obtain instructions for which the cache behavior differs between two versions. However, doing so would depend heavily on cache placement and replacement policies, and on the cache size and layout. 

\fakepar{Peripherals} Instructions that enable interactions with external peripherals, such as sensors using serial interfaces or motors using pulse-width modulation, are inherently difficult to reason about in time.
\deltatool can be combined with time measurements obtained by profiling the actual executions on hardware to support these configurations, with a few caveats.

First, one needs to isolate the relevant code blocks and make sure the time to execute the block must  either be independent of the rest of the program, or the dependency must be known.
In case there is a dependency, all possible execution times for the code block must be measured to establish an execution time interval.
Additionally, the effects of the code block on the domains of the processor's registers after its execution must be known to guarantee that it does not affect timing elsewhere.

If these conditions are met, then the code block can be replaced with a single artificial instruction with the measured execution time(s).
This instruction should replicate any change to register domains that the actual execution of the relevant code block would perform.
For example, a code block that reads the value of the sensor and writes it to a register should be replaced with an artificial instruction that writes a value within a specific interval to the same register. 


\section{Evaluation} \label{sec:evaluation}

Our evaluation is three-pronged.
\secref{benchmarks} introduces the benchmarks we consider.
\secref{evaluation_execution_time} compares the execution time differences computed using \deltatool with those measured on real hardware, providing an assessment of \emph{absolute} accuracy of the differential timing estimates.
\secref{worst_case_execution_times} compares WCET estimates of updated programs using \deltatool and \ait~\cite{ait}, offering a \emph{relative} accuracy assessment against existing tools.
In \secref{evaluation_analysis_time}, we investigate the improvements in computational effort due to using \reta in either \deltatool or \aitreta. 
We conclude the evaluation in \secref{scalability} with a note on scalability.
Our results lead to five key conclusions:
\begin{enumerate}
\item in all benchmarks but one, the estimates returned by \deltatool are safe compared to real hardware executions, and oftentimes quite close to the latter in absolute value;
  \item WCET information returned by \deltatool ranges from \emph{exactly} the WCET observed on real hardware to 148\% of the new version's measured WCET;
  \item in the same benchmark as point 2), aiT measures a  12\% and 149\% higher WCET: \deltatool is thus either more accurate or similarly pessimistic compared with aiT;
  \item the (partial) implementation of \reta in aiT reduces the analysis time and memory consumption by 45\% and 8.9\%, respectively;
  \item removing \reta from \deltatool, effectively reducing the latter to operate as a regular timing analysis tool, increases its computational complexity by 27\%.
\end{enumerate}

\subsection{Benchmarks} \label{benchmarks}

We analyze five programs.
We choose the programs as key examples of embedded software and to test \reta against different code structures and various updates.

\fakepar{Matrix multiplication} Matrix multiplication is a common operation in image analysis, digital signal processing, and control systems~\cite{woehrle_caemo_2018}, while serving as a base micro-benchmark for our evaluation. 
We consider two separate updates, both marginally impacting the source code.
One update adds an operation in the multiplication procedure: scaling the matrix elements by a factor of 2.
The other update increases the sizes of the input matrices from 32x32 to 64x64.

\fakepar{Sorting} We update a sort program that sorts a list of 5000 elements, changing from bubble sort to insertion sort.
In this case, the update changes most of the original code.
The execution time of a sorting algorithm depends on the size and ordering of the input list.
Bubble sort and insertion sort have their worst-case execution time when the list is sorted in the reverse order, and the best-case execution time when the list is already sorted in the correct order~\cite{drozdek_2013}.
We compute the execution time difference when updating the program from bubble sort to insertion sort both when the input is a reverse sorted list and an already sorted list.

Updating the sorting algorithm is an update where functional equivalence between the two versions is retained, that is, the two versions have the same input-output relations, but where the implementations are different. Performance optimizations that retain functional equivalence are common in resource-constrained devices.

\fakepar{Proximity detection} We implement a proximity detection functionality, as used in mobile robotics~\cite{9636130}, that uses a time-of-flight ranging sensor (VL53L1X~\cite{vl53l1x}) to manipulate an LED based on the distance from the sensor to the nearest object.
The original version of the program turns an LED on if the object is within 50 centimeters, and off otherwise.
The updated version brightens the LED as the object moves closer and darkens it as the object moves further.
We compute the execution time differences between the two versions of the program to find the timing impact of incorporating finer LED control, thus modifying about 13\% of the original source code.

\fakepar{Fast Fourier transform} We use ArduinoFFT~\cite{arduinoFFT}, an existing fast Fourier transform implementation, to estimate the frequency of a simulated 1000 Hz fictitious signal.
To utilize the FPU on the target MCU, we adjust the library beforehand to use single-precision floating point numbers instead of double.
Increasing the number of samples used to compute the Fourier representation improves accuracy but increases the execution time.
The original version of the program uses 64 samples, whereas the updated version uses 1024 samples.
This requires changes to about 1.3\% of the original source code.

\fakepar{Hackflight} Hackflight is an existing low-level flight controller for aerial drones~\cite{levy_robustness_2020}.
We consider an actual update performed on the Hackflight codebase that adjusts the trim for the yaw inputs~\cite{hackflight_yaw_update}, retrieved from the Hackflight GitHub repository. 
It is a realistic update within the embedded software domain.
Yaw represents the angle at which the drone is pointing towards.
The trim is a fixed value added to the controller's set-point to shift its value by a specific offset, and is commonly used to counteract drone drift.
To facilitate the analysis, we use constant sensor inputs. 

\subsection{Accuracy} \label{evaluation_execution_time}

We determine the accuracy of the execution time difference computed with \deltatool compared with real hardware.

\fakepar{Setup}
We use \deltatool to determine the execution time differences between the two program versions.
Next, we run each program version on a NUCLEO-L432KC development board~\cite{nucleo-l432kc}, equipped with an ARM M4F MCU, and measure the execution time using the Data Watchpoint and Trace (DWT) module.
This module is included in some ARM Cortex-M MCU and counts the number of execution cycles.
By reading the count prior to, and after a section of code we determine how many cycles that section of code required.
We capture 500 samples for each benchmark.

\begin{table*}[]
  \centering
  \caption{Execution time differences for various benchmarks obtained through hardware execution and \deltatool.}
  \label{tab:execution_time_differences_results}
  \begin{tabular}{|l|c|c|}
  \hline
  \multicolumn{1}{|c|}{\textbf{Updates}} & \textbf{Hardware Execution (Cycles)} & \textbf{\deltatool (Cycles)} \\ \hline
  Adding a 2x scaling factor to 32x32 matrix multiplication & 32765 to 32769 & 32768 \\ \hline
  Doubling the inputs of matrix multiplication & 1,896,974 to 1,896,989 & 1,731,744 to 2,926,944 \\ \hline
  Switching from bubble sort to insertion sort on an already sorted array & 29970 to 30093 & 4991 to 59983 \\ \hline
  Switching from bubble sort to insertion sort on a reverse sorted array & -162,596,596 to -162,595,489 & -324,975,018 to -149,980,015 \\ \hline
  Proximity detector adjusting LED's brightness & 1552 to 1579 & 1534 to 1588 \\ \hline
  Changing the FFT's samples from 64 to 1024 & 398,068 to 399,488 & 355,495 to 410,381 \\ \hline
  Adjusting trim for yaw inputs in Hackflight & 0 & -24 to 24 \\ \hline
  \end{tabular}
  \end{table*}

\begin{figure*}[t]
  \begin{subfigure}[b]{.24\textwidth}
    \centering
  \includegraphics[width=\textwidth]{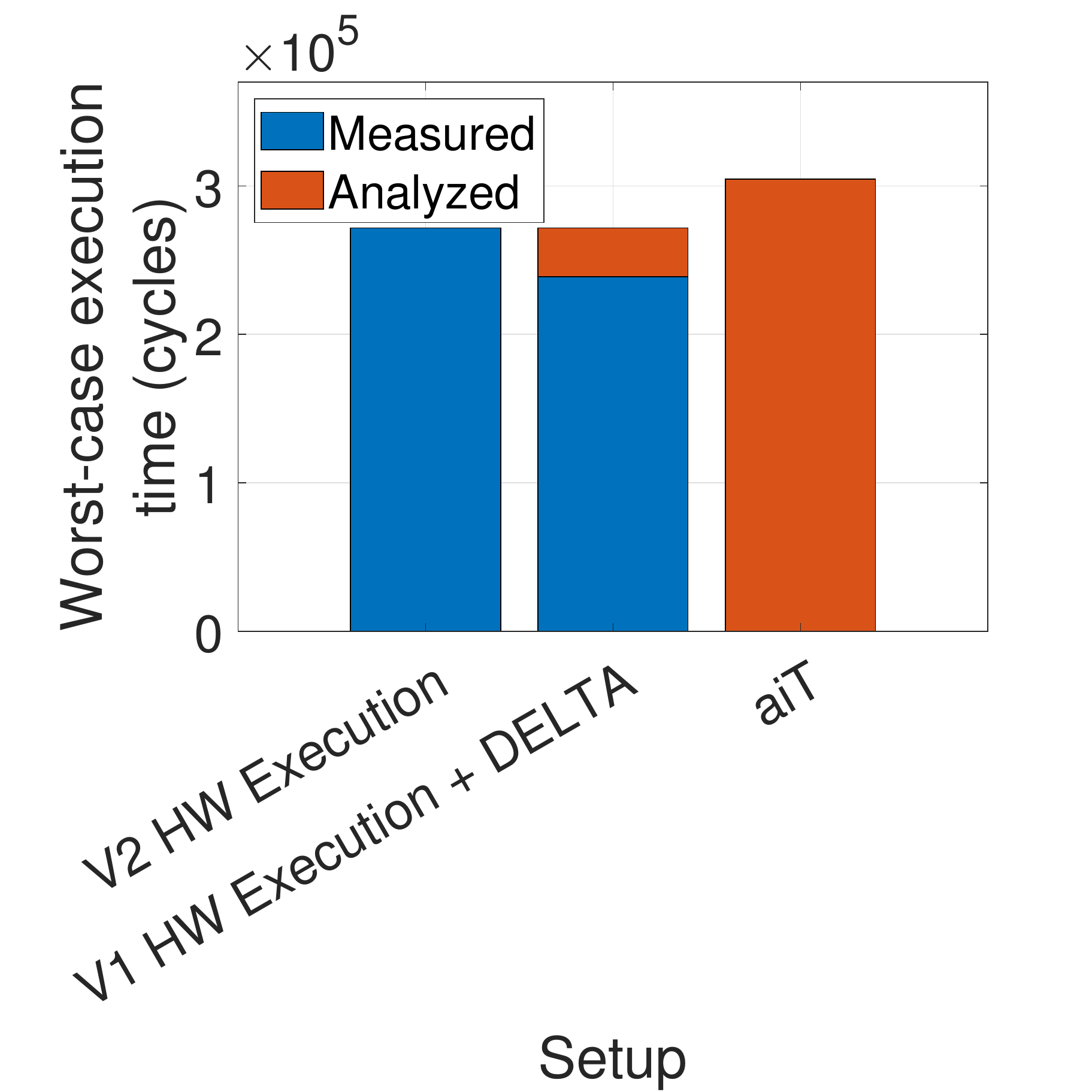}  
    \caption{Matrix multiplication and 2x \\ scaling of two 32x32 matrices.}
    \label{fig:matmult_2x_wcet}
  \end{subfigure}
  \hfill
  \begin{subfigure}[b]{.24\textwidth}
    \centering
    \includegraphics[width=\textwidth]{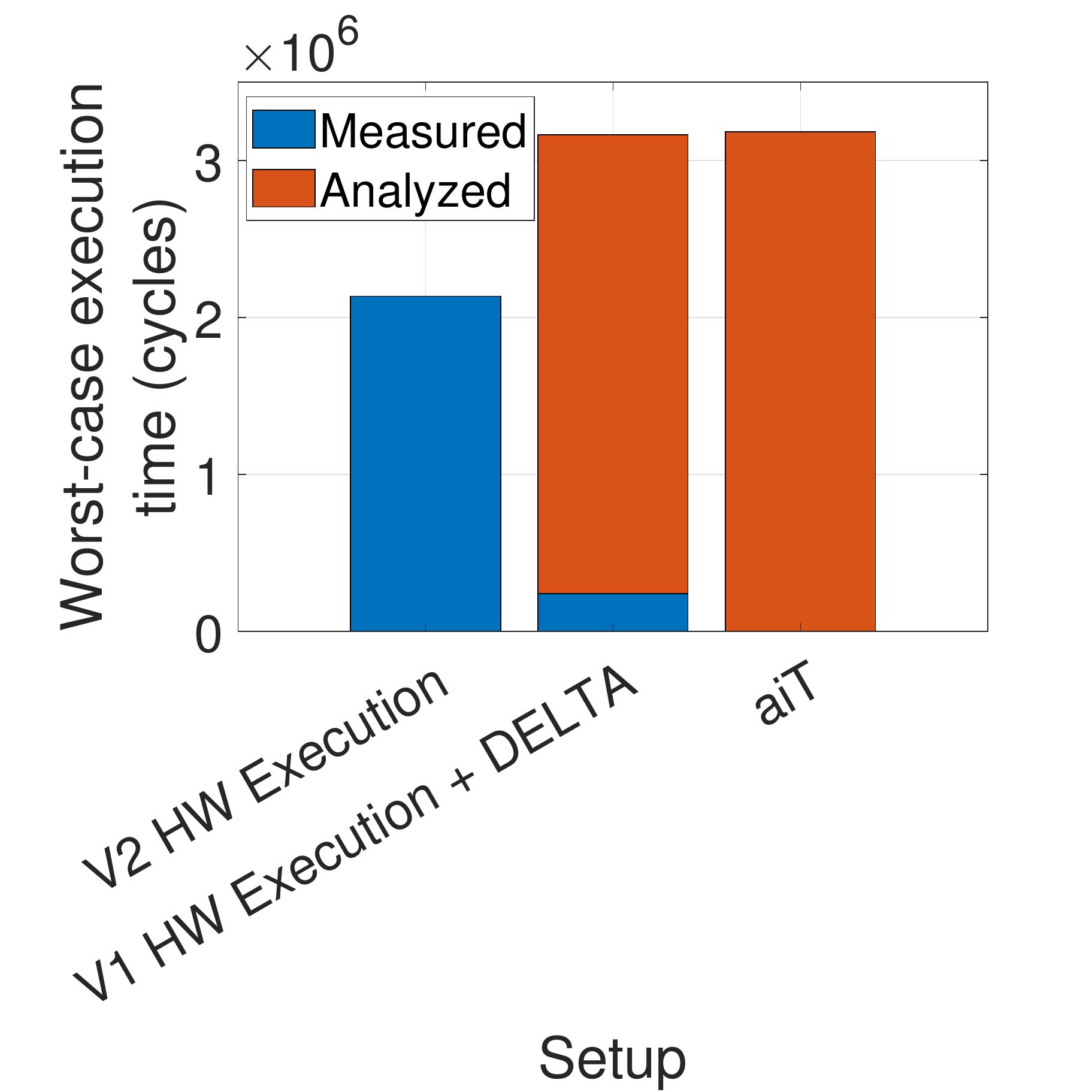}  
    \caption{Matrix multiplication of \\ two 64x64 matrices.}
    \label{fig:matmult_64_wcet}
  \end{subfigure}
  \hfill
  \begin{subfigure}[b]{.24\textwidth}
    \centering
  \includegraphics[width=\textwidth]{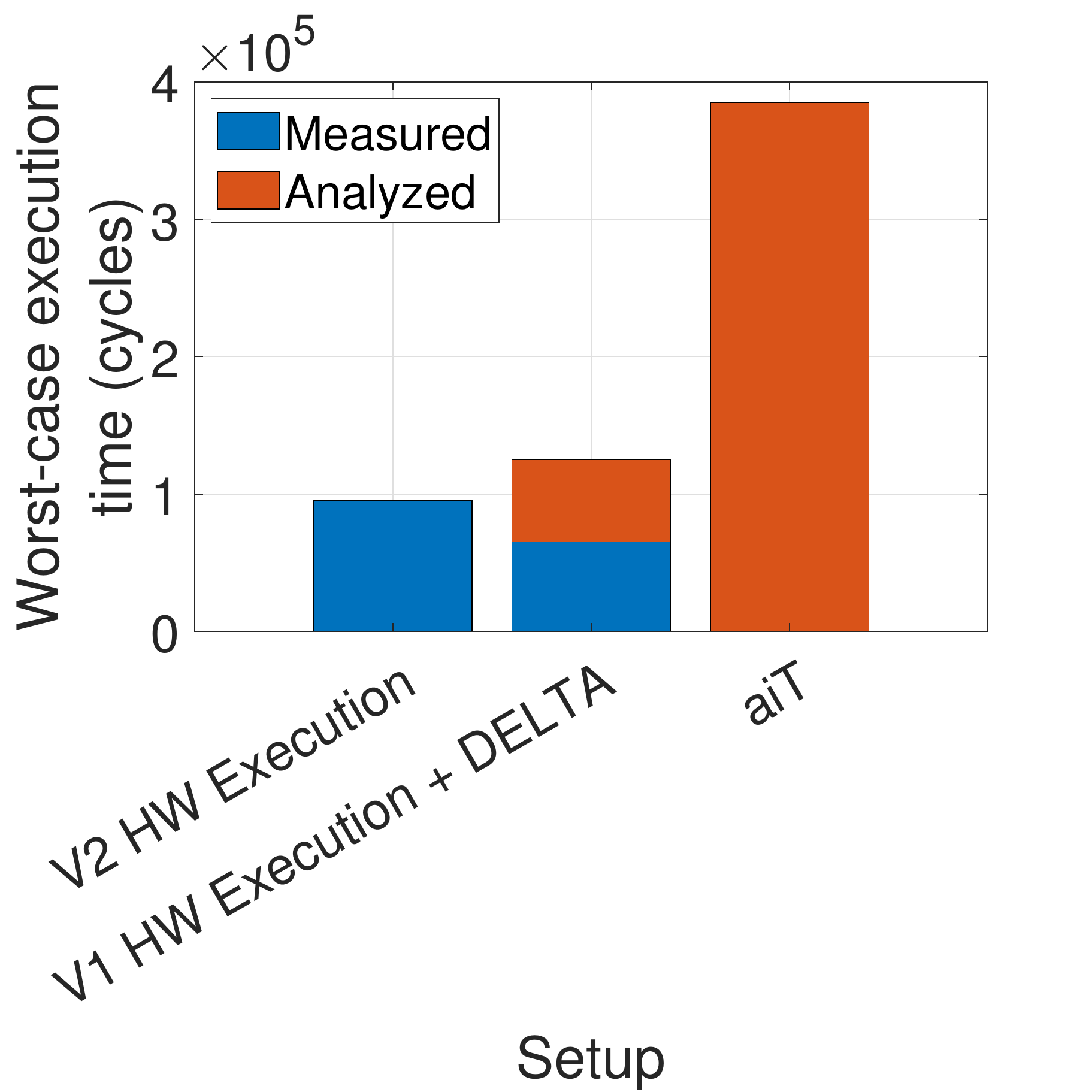}  
    \caption{Insertion sort of an already \\ sorted array of 5000 elements.}
    \label{fig:sorted_array_wcet}
  \end{subfigure}
  \hfill
  \begin{subfigure}[b]{.24\textwidth}
    \centering
    \includegraphics[width=\textwidth]{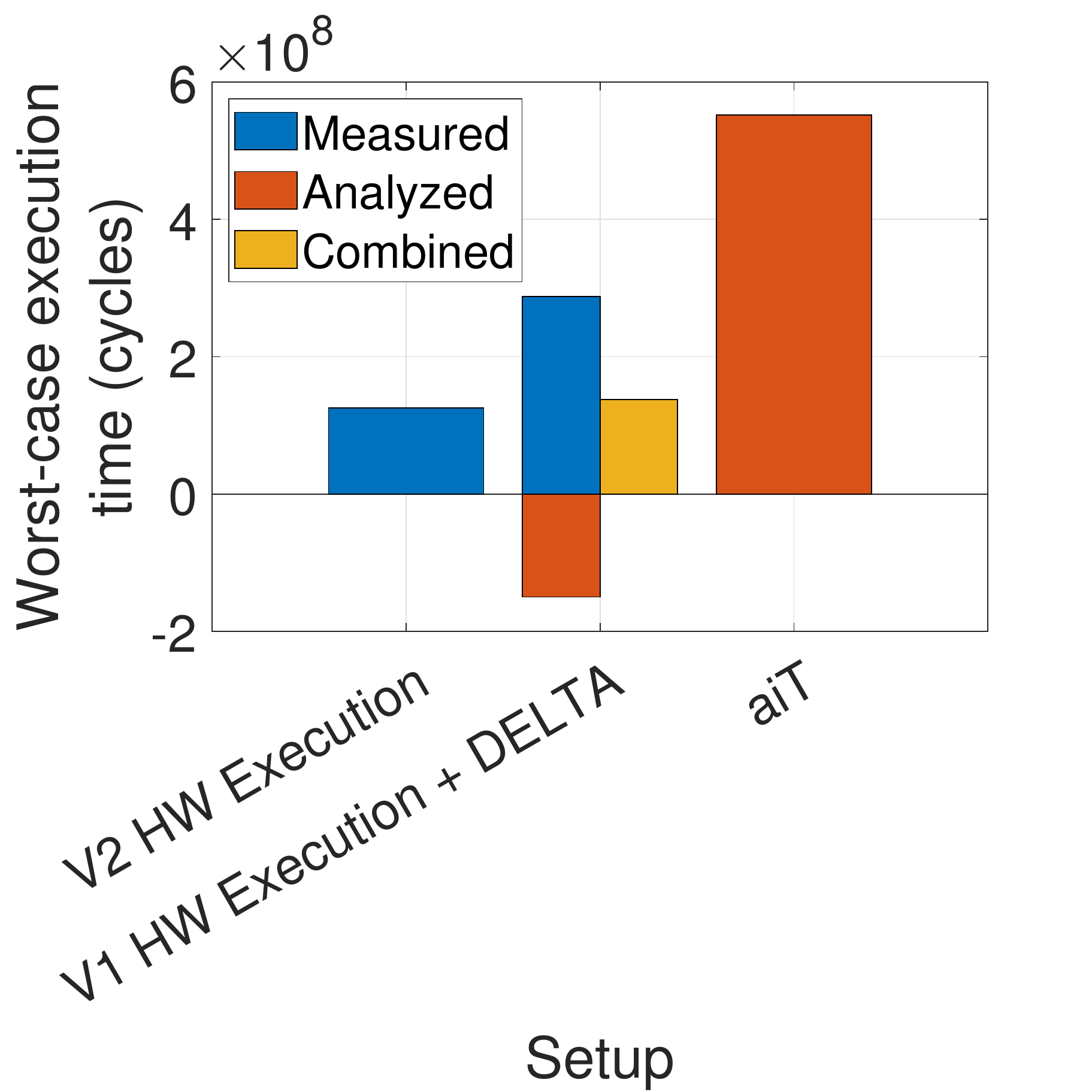}  
    \caption{Insertion sort of a reverse \\ sorted array of 5000 elements.}
    \label{fig:reverse_sorted_array_wcet}
  \end{subfigure}
  \begin{subfigure}[b]{.32\textwidth}
    \centering
    \includegraphics[width=\textwidth]{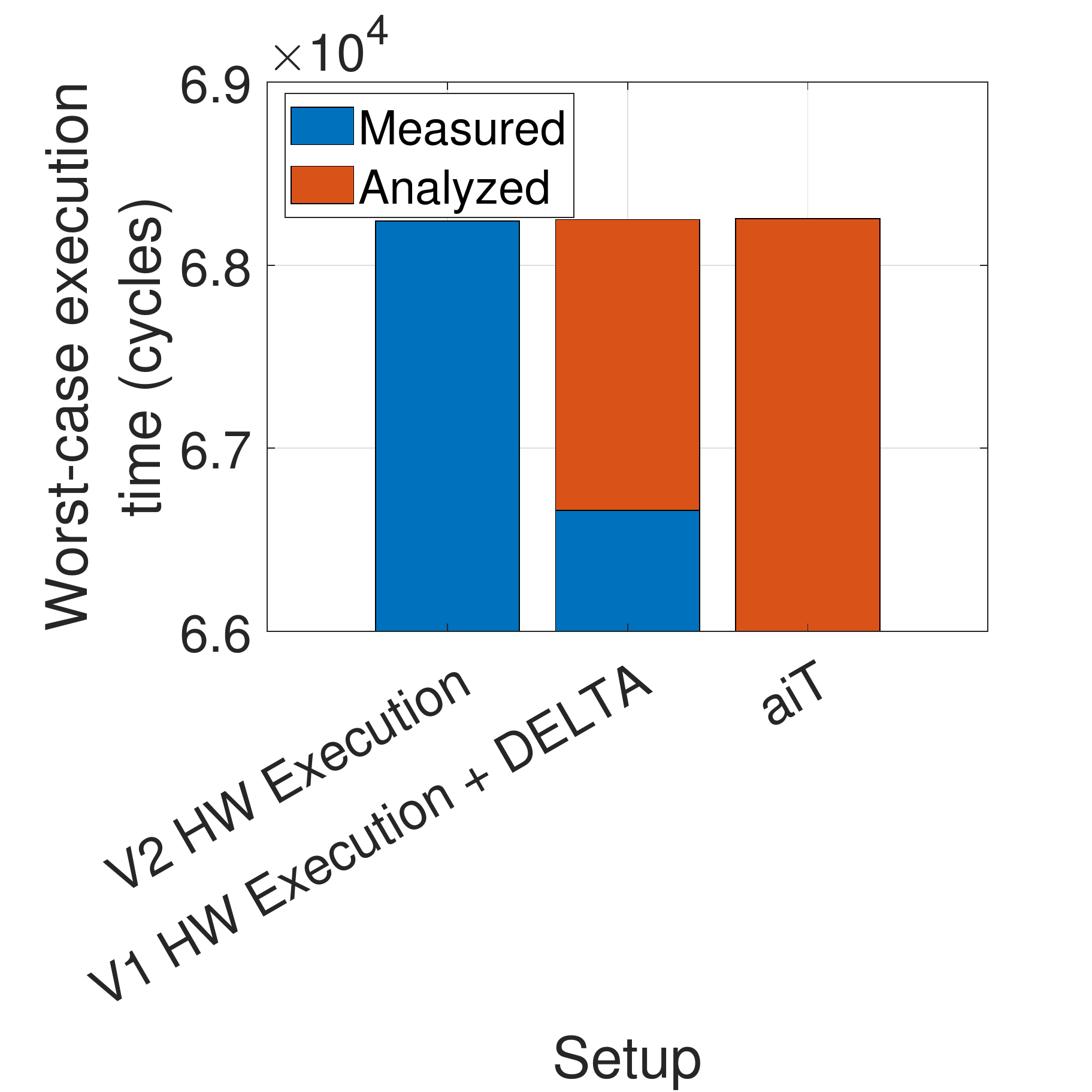}  
    \caption{Adjusting LED brightness depending \\ on distance from ranging sensor.}
    \label{fig:rangefinder_fade_wcet}
  \end{subfigure}
  \begin{subfigure}[b]{.32\textwidth}
    \centering
    \includegraphics[width=\textwidth]{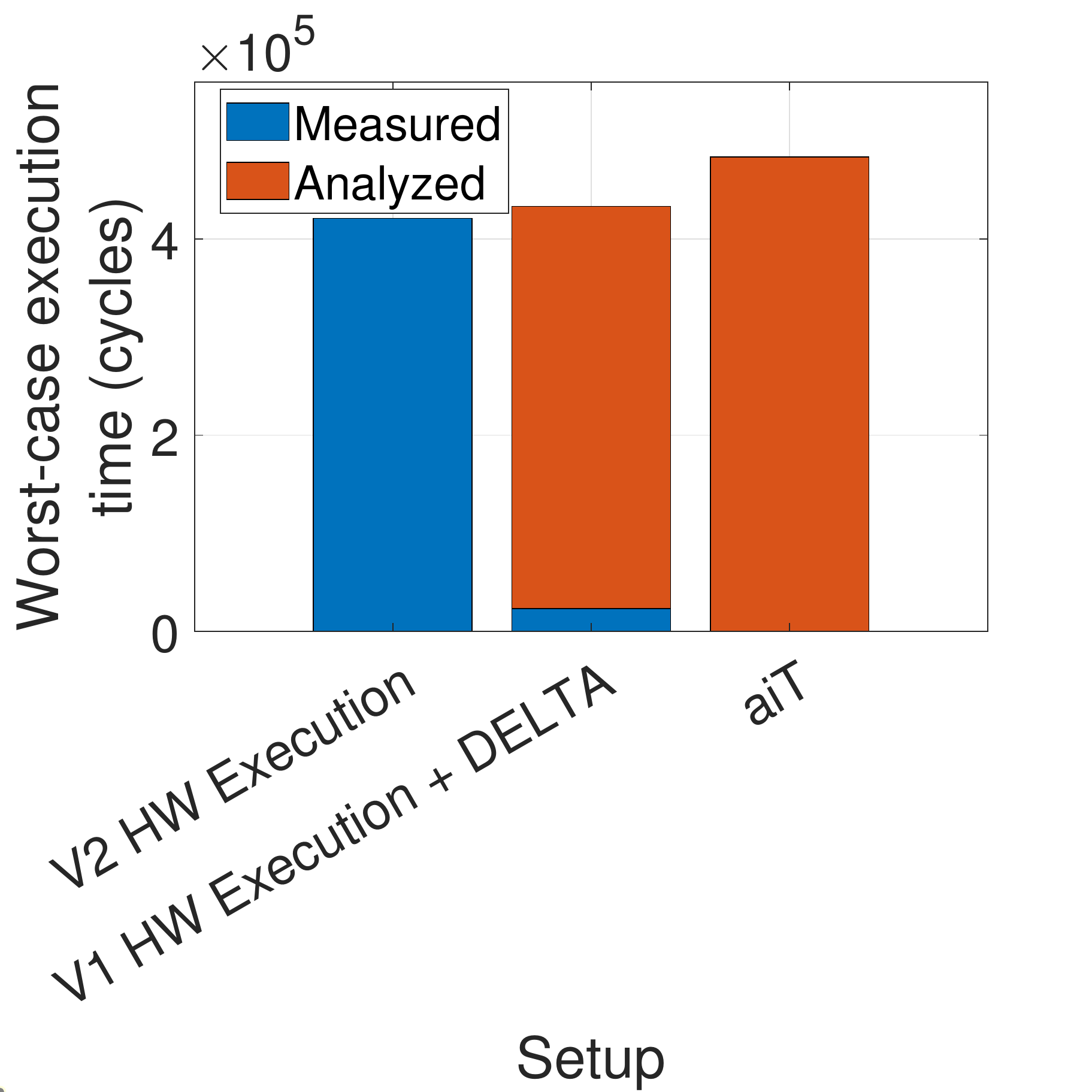}  
    \caption{Performing an FFT of a sine wave \\ to find its frequency using 1024 samples.}
    \label{fig:example_7_FFT_1024_wcet}
  \end{subfigure}
  \begin{subfigure}[b]{.32\textwidth}
    \centering
    \includegraphics[width=\textwidth]{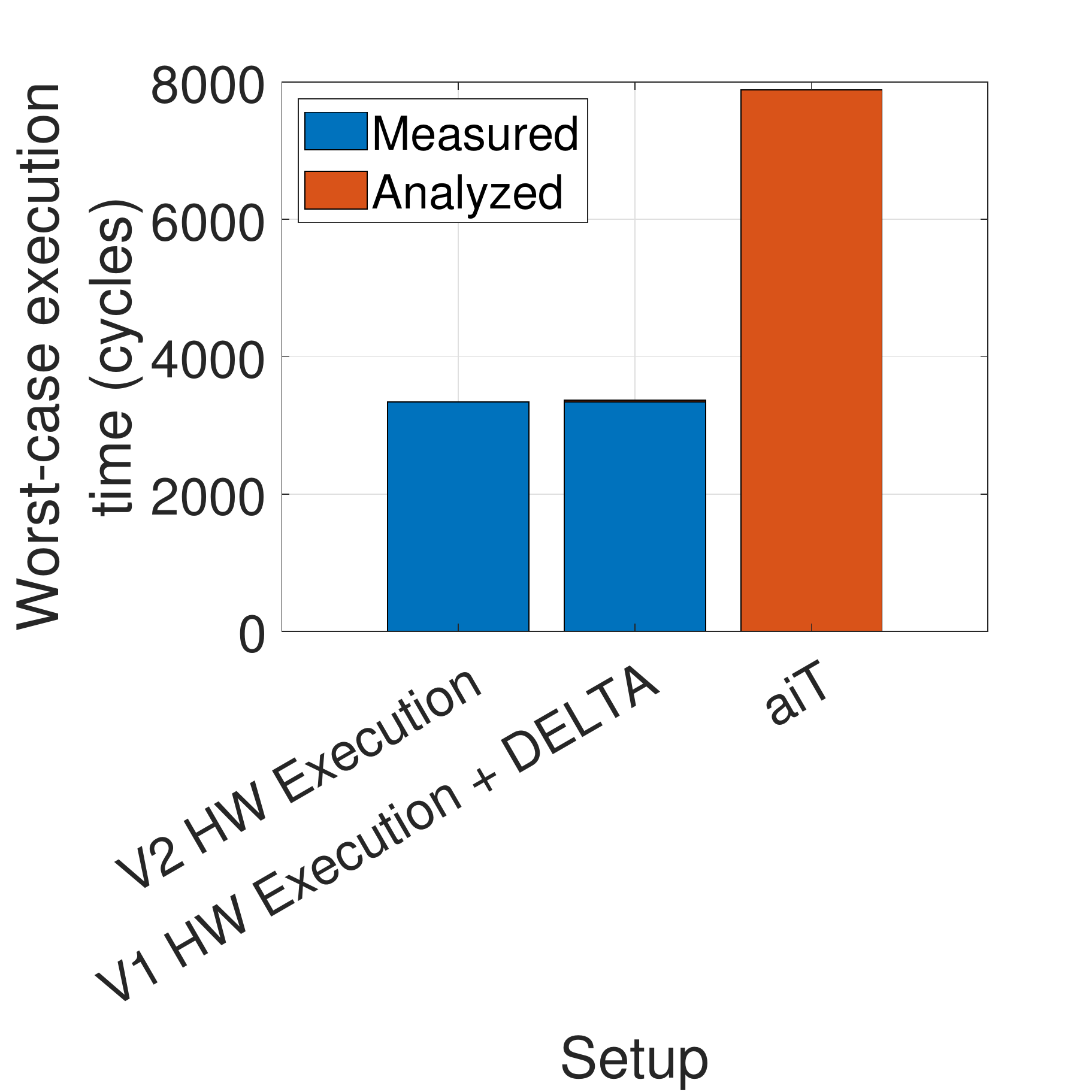}  
    \caption{Hackflight's control loop in the \\ disarmed state.}
    \label{fig:hackflight_wcet}
  \end{subfigure}
  \caption{Worst-case execution times for various benchmarks obtained through three techniques.}
  \label{fig:wcet_results}
\end{figure*}

\fakepar{Results} \tabref{tab:execution_time_differences_results} shows the results.
Hardware execution figures are intervals because we observe different execution times across the 500 measurements.
\deltatool's results are intervals because variations in the execution time of instructions introduce the need for pessimistic and optimistic execution time differences, as explained in \secref{analysis_imprecision}. 


For \deltatool to provide a safe estimate of the execution time difference, all possible hardware execution times must fall within the \deltatool estimated interval.
The update that adds a scaling factor \emph{is the only case where this does not happen}.
This is because the optimistic and pessimistic differences computed by \deltatool are identical; however, the time observed on real hardware varies between four cycles.
As explained in \secref{analysis_imprecision}, our analysis excludes cache and pipeline effects, which may result in varying execution times.
A variation of four cycles out of 30+K cycles is less than 0.02\%, and hence bears a very limited impact overall.
Instead, when doubling the matrix dimensions, all possible hardware execution times are comprised within the \deltatool estimated interval.


As for the sorting benchmark, the hardware execution times completely fall within \deltatool's estimates.
The latter provide further useful insights, such as determining that moving from bubble sort to insertion sort is guaranteed to improve performance for reverse sorted arrays and guaranteed to worsen performance for already sorted arrays.
It also provides definitive bounds on the performance degradation.
Based on this, a developer may decide that the improvement in execution time for the reverse sorted case, which is guaranteed to be of at least 149 million cycles, is worth the performance degradation for sorted arrays, which is guaranteed to be at most 59983 cycles.

Hardware execution times for the other three benchmarks: proximity detection, FFT, and Hackflight also completely fall within \deltatool's estimates, making the output of the latter a safe approximation.
Note how applying \deltatool for the proximity detection and FFT benchmarks involves the technique explained in \secref{analysis_imprecision} to integrate hardware measurements.
In the former, we experimentally measure the time to read from a sensor and to execute the \textit{AnalogWrite} routine.
In the latter, we experimentally measure the time to execute the \textit{cosf} routine for every possible input used in the benchmark because explicitly analyzing this routine would require intimate knowledge on recursive numerical algorithms.

\subsection{Worst-case Execution Times} \label{worst_case_execution_times}

We test \deltatool when estimating WCET after a code update, compared with ground truth information obtained from real hardware executions and WCET estimations provided by \ait.


\fakepar{Setup} As \deltatool does not provide absolute WCET information, we proceed in three steps. 
First, we measure the largest observed execution time on real hardware \emph{for the new version}, with the same setup as \secref{evaluation_execution_time}.
This represents ground truth information.
Second, we measure the largest observed execution time on the hardware for the original version; we use this as a baseline to add the pessimistic execution time difference computed by \deltatool, ultimately computing the WCET of the new version.
This is how a developer would normally operate when using \deltatool.
Third, we obtain the WCET using \ait, that is, how a developer would operate in the absence of \deltatool.

\fakepar{Results} \figref{fig:wcet_results} shows the results.
In one of the matrix multiplication benchmarks, depicted in \figref{fig:matmult_2x_wcet},  the WCET computed by \deltatool is very close to the new version's measured execution time.
The WCET obtained by \ait is way more (and unnecessarily) pessimistic.
\figref{fig:matmult_64_wcet}, instead showing the results for the other matrix multiplication benchmark, indicates that the WCET computed by \deltatool is almost the same as the one computed through \ait.
They are both more pessimistic than the new version's measured execution time.
\tabref{tab:execution_time_differences_results} demonstrates that the range of possible execution times computed with \deltatool is large in this case.
As a result, we see a larger degree of pessimism in the WCET information we obtain from \deltatool when compared with \figref{fig:matmult_2x_wcet}.
Regardless, \deltatool has a similar degree of pessimism as \ait.

The results of the sorting benchmarks are shown in \figref{fig:sorted_array_wcet} and \figref{fig:reverse_sorted_array_wcet}. \figref{fig:reverse_sorted_array_wcet} differs from prior figures in that the execution time difference computed by \deltatool is negative.
This means that the new versions takes less time than the old version.
Therefore, \figref{fig:reverse_sorted_array_wcet} includes an additional bar in yellow to indicate the sum of the measured execution of the original version (blue bar) and the negative execution time difference (orange bar).
The yellow bar is the WCET of the new version.
Both charts demonstrate that the WCET obtained by \deltatool is close to the measured WCET of the updated version.
\ait is again more (and unnecessarily) pessimistic.
This is due to \ait assuming the longest path is always taken in each loop iteration in the sorting algorithms, whereas \deltatool uses information about when and how often a specific path is taken given the order of the input list. 

The results of the proximity detection benchmark is shown in \figref{fig:rangefinder_fade_wcet}. \ait's result involves a measurement-based component due to the measurement of the \textit{SensorRead} and \textit{AnalogWrite} functions. The latter is incorporated into \deltatool's difference from \tabref{tab:execution_time_differences_results} as mentioned earlier.
All three WCETs are close to each other.
In comparison with the measured one, \deltatool's and \ait's are 7 and 13 cycles larger respectively.

Finally, Both \figref{fig:example_7_FFT_1024_wcet} and \figref{fig:hackflight_wcet} demonstrate that \deltatool's WCET estimates are closer to the hardware execution's than to \ait. 
In particular, Hackflight includes a loop that cycles through the PID controllers in sequence.
\ait assumes that the most time-consuming controller is run in each iteration of the loop, which contributes to a very pessimistic WCET estimate.
As for FFT, the difference between \ait and \deltatool is likely caused by the overly pessimistic choices of the paths taken when estimating the WCET.

\subsection{Computational Effort} \label{evaluation_analysis_time}

To complete our evaluation, we investigate the impact of differential timing analysis on computational effort, comparing \reta with traditional timing analysis.

\fakepar{Setup} We provide \emph{three different views} on computational effort.
First, we compare the \emph{processing time} and \emph{peak memory consumption} incurred by the regular \ait with that of \aitreta, that is, a version of \ait we customize to use parts of the differential timing analysis technique of \reta.
We recall that as the implementation of \aitreta is only partial, it cannot process arbitrary updates.

We use an Intel Core i7-1065G7 running version 22.10 of \ait on Windows 11 Pro for these experiments. 

We also compare the \emph{computational complexity} of \deltatool with and without \reta.
As hinted earlier, removing \reta from \deltatool effectively makes the latter operate as a traditional timing analysis tool.
Based on knowledge of the different steps in \reta and the number of times a given piece of information is used in the analysis, we estimate the computational complexity of \deltatool as
\begin{equation} \label{eq:computational_complexity_diff}
CC = 4*(A+B)+3*D+3*F+5*G,
\end{equation} where A is the number of instructions inside loops whose loop bounds changed, B is the number of removed/added instructions inside loops whose loop bounds remained the same, D is the number of removed/added instructions outside loops, F is the the number of lines included in the forward slice, and G is the the number of lines included in the backward slice.


Differently, when we remove \reta from \deltatool, hence the entire program is analyzed regardless, we estimate the computational complexity as
\begin{equation} \label{eq:computational_complexity_full}
CC = 4*X+3*Y+5*Z, 
\end{equation} where X refers to the number of instructions inside loops, Y refers to the number of instructions outside loops, and Z refers to the lines included in the backward slice of all conditional statements, which is used anyways for dataflow analysis. 

To make \aitreta applicable, we analyze a program that originally contains a single code block performing the multiplication of two 32x32 matrices.
In a first test, the added code block multiplies two other matrices and scales the result matrix's values by 2x.
In the second test, the added code block just scales a matrix's values by 2x.




\begin{figure}[t]
  \begin{subfigure}[b]{.24\textwidth}
    \centering
    \includegraphics[width=\textwidth]{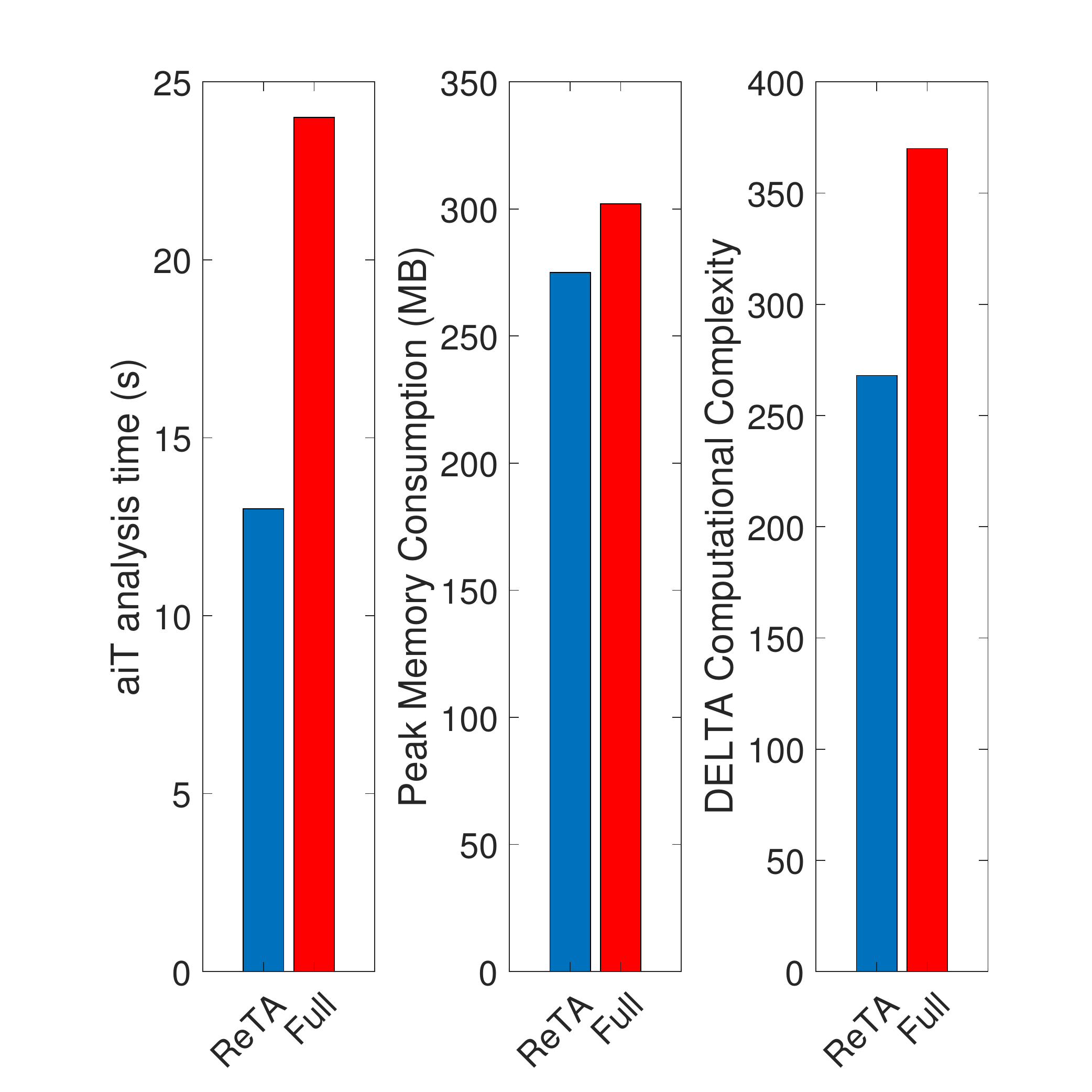}  
    \caption{Adding matrix multiplication and 2x scaling.}
    \label{fig:example_10_analysis_time}
  \end{subfigure}
  \hfill
  \begin{subfigure}[b]{.24\textwidth}
    \centering
    \includegraphics[width=\textwidth]{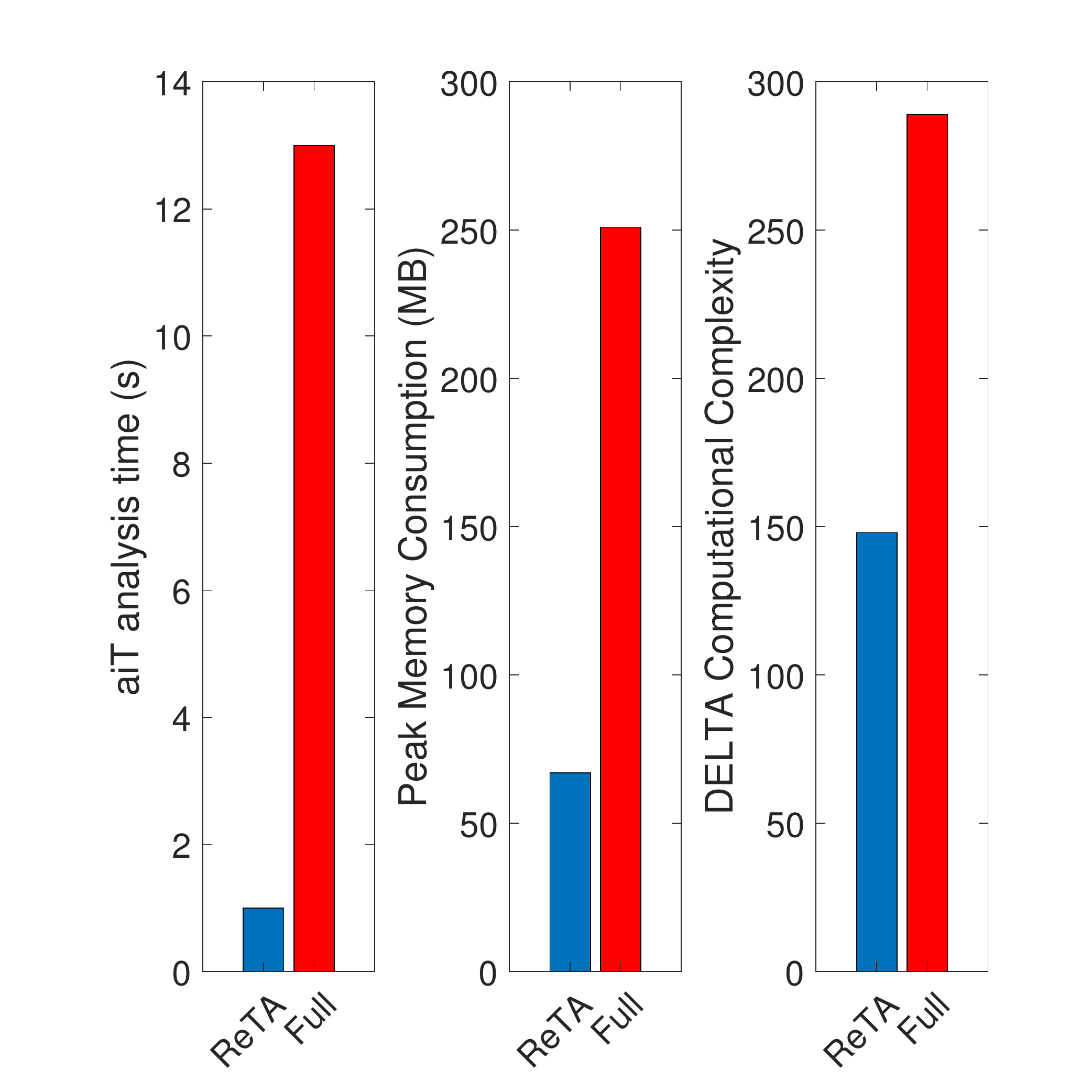}  
    \caption{Adding 2x scaling factor.\\~}
    \label{fig:example_11_analysis_time}
  \end{subfigure}
  \caption{Computational complexity.}
  \label{fig:analysis_time}
\end{figure}

\fakepar{Results} \figref{fig:analysis_time} summarizes the results.
\figref{fig:example_10_analysis_time} demonstrates that in comparison with \ait, \aitreta's analysis time and peak memory consumption decrease by 45\% and 8.9\% respectively.
Employing \reta in \deltatool also results in a 27\% reduction of the latter's computational complexity.
In the case of \figref{fig:example_11_analysis_time}, the added code block is smaller than the existing code block.
\aitreta's analysis time and peak memory consumption decrease by 92\% and 76\% respectively.
The use of \reta in \deltatool also results in a 49\% reduction in its complexity.


It is interesting to observe that the improvement due to the use of \reta is generally smaller in \deltatool than \aitreta.
This is caused by the extra overhead of the slicing procedure, not used in \aitreta.
This overhead partly overweighs the benefits of differential timing analysis, yet is necessary for \deltatool to be able to analyze a significantly larger set of updates than \aitreta.

\subsection{Scalability} \label{scalability}
A prior study of 43 programs demonstrates that backward and forward slices take an average of 0.18 seconds for programs with 23,421 lines of code, on average~\cite{1235405}. Even for the most complex program in the study, which has 93,309 lines of code, around 6 million vertices, and 29 million edges in its CFG, the average time per slice was 2.7 seconds.
Determining whether an instruction belongs to CatA or CatB is trivial.
Differentiating between CatB\textsubscript{a} and CatB\textsubscript{b} is more challenging since we must identify whether the instruction leads to a natural loop.
A loop detection algorithm that takes the CFG as input has approximately linear time complexity~\cite{wei2007new}. 
Building the CFG from assembly code also has linear time complexity provided all branches have known targets~\cite{cooper2002building}, which is a requirement we specify in \secref{reta_requirements}.


The slicing procedure does not return the minimal program enabling timing analysis. Computing this program requires additional iterations, impacting scalability.
We choose the specific slicing procedure as a tradeoff between the size of the resulting program and the computational complexity required to determine the slice.


\section{Conclusion}

We presented \reta, a differential timing analysis technique that uses program slicing, statement categorization, and dataflow analysis to estimate the impact of an update on the execution time of embedded software.
We developed an implementation of \reta called \deltatool for ARM M4 MCUs and customized the existing \ait tool to incorporate a subset of the \reta procedures.
We demonstrated that the timing estimates provided by \deltatool are oftentimes in line with actual measurements from real hardware and that the WCET estimates of \deltatool range from exactly the WCET of real hardware to 148\% of the new version's measured WCET. In the same setting, the unmodified
\ait estimates are 112\% and 149\% of the actual executions; therefore, \deltatool is either more accurate or as pessimist as an industry-strength tool such as \ait, while requiring a much lower computational effort.


\section*{Acknowledgment}
Work partially funded by the Knut and Alice Wallenberg Foundation through project UPDATE.
This work is supported by the Swedish Foundation for Strategic Research (SSF).
\bibliographystyle{IEEEtran}
\bibliography{references}

\end{document}